\documentclass[10pt]{article}

\usepackage{epsfig}
\usepackage{subfigure}
\usepackage{algorithm, algorithmic}
\usepackage{amssymb}

\newtheorem{theorem}{\textsc{Theorem}}
\newtheorem{corollary}{\textsc{Corollary}}
\newtheorem{definition}{\textsc{Definition}}
\newtheorem{example}{\textsc{Example}}
\newtheorem{proof}{\textsc{Proof}}

\begin{document}
\title{Analyzing and Improving Performance of a Class of
Anomaly-based Intrusion Detectors}

\author{Zhuowei Li, Amitabha Das \thanks{Center for Advanced Information Systems, School of
Computer Engineering, Nanyang Technological University, 50,
Nanyang Avenue, Singapore, 639798. \textbf{Email}:
zhwei.li@pmail.ntu.edu.sg, asadas@ntu.edu.sg}}

\maketitle
\begin{abstract}
Anomaly-based intrusion detection (AID) techniques are useful for
detecting novel intrusions into computing resources. One of the
most successful AID detectors proposed to date is stide, which is
based on analysis of system call sequences. In this paper, we
present a detailed formal framework to analyze, understand and
improve the performance of stide and similar AID techniques.
Several important properties of stide-like detectors are
established through formal proofs, and validated by carefully
conducted experiments using test datasets. Finally, the framework
is utilized to design two applications to improve the cost and
performance of stide-like detectors which are based on sequence
analysis. The first application reduces the cost of developing AID
detectors by identifying the critical sections in the training
dataset, and the second application identifies the intrusion
context in the intrusive dataset, that helps to fine-tune the
detectors. Such fine-tuning in turn helps to improve detection
rate and reduce false alarm rate, thereby increasing the
effectiveness and efficiency of the intrusion detectors.
\end{abstract}


\section{Introduction}
Since the concept of intrusion detection in computer systems was
proposed by Anderson \cite{Anderson80IDS}, many research studies
have been carried out to find appropriate intrusion detection
techniques to protect the resources in computers or networks
\cite{Neumann99ExperienceEMERALD} \cite{Lee00FrameworkMADAM}
\cite{Mahoney02LERADPaper} \cite{Calvin97ExeMonitoring}. However,
the network disaster caused recently by Nimda, MSBlast and
MSSasser highlights the shortcomings of the intrusion detection
techniques deployed in our network infrastructures
\cite{Kemmerer02IntrusionHistory}, and indicates that intrusion
detection techniques still have a long way to go before they can
provide effective protection to computing resources.

In general, the intrusion detection techniques can be categorized
into \textit{signature-based intrusion detection (SID)} and
\textit{anomaly-based intrusion detection (AID)} ones. The SID
techniques  build (and/or update) intrusion signature bases that
include the signature of every known intrusion. Then, the resource
behavior that matches any intrusion signature in the bases is
labeled as an intrusion. Obviously, previously unknown intrusions
cannot be detected by this technique. Besides this drawback, the
requirement of instant updating of the intrusion signature bases
imposes a severe performance bottleneck on SID techniques.

Anomaly-based intrusion detection techniques have become a focus
of intense research as they offer an useful alternative to SID,
that is capable of detecting novel intrusions. One of their
implicit assumptions is that the violations or anomalies are
indications of intrusions, i.e. the anomaly is caused by an
intrusion into the resource. This assumption, though largely
correct, may not be always true as some malicious intrusions may
not violate the parameters of normal behaviors, whereas some
non-malicious activities may appear as violations
\cite{Tan02HidingIntrusion}.

Almost all AID techniques work as follows. First, a model of
normally behaving users \cite{JavitsNIDESStatisticalComponent}
\cite{Ju01HighOrderMarkov} and/or processes
\cite{Hofmeyr98SeqSystemCall} \cite{Mahoney02LERADPaper} is built.
Then, an intrusion is detected by comparing the actual current
behavior against the normal model and taking actions according to
some predetermined security policies \cite{Calvin97ExeMonitoring}
\cite{Chari03BlueBox} \cite{Sekar02SpecAD}. Although many AID
techniques have been proposed to date, no single AID technique can
effectively detect all types of intrusions into the resources
under various scenarios \cite{Kemmerer02IntrusionHistory}. More
specifically, they suffer from high false alarm rate that tends to
reduce the effectiveness of true alarms because of the base rate
fallacy \cite{Axelsson99BaseRateFallacy}. In addition, the high
cost of and insufficient guideline about training the normal model
do not make matters any better for AID.

In this paper, instead of proposing a new AID technique, we
develop a formal framework to argue about and analyze the
properties of one typical AID technique called \textit{stide}
\cite{Forrest94NSA} \cite{Forrest96SenseSelf}. Generally speaking,
to be efficient, any AID technique must try to increase the
detection rate simultaneously keeping the false alarm rate to a
minimum. For this reason, it is important to understand the
factors that suppress the detection rate, and lead to false
alarms. In this framework for stide, the factors are identified as
the minimum foreign sequences in the intrusive dataset and the
maximum self sequences in the test dataset, and the relations
between these factors and stide efficiency are expressed and
discussed. In addition, most related works to stide are
interpreted in a logical way under this framework, namely, mimicry
attacks, information hiding techniques, t-stide, variable-length
patterns, and locality frame scheme.

Our aim in this paper is just that, and we do so by providing a
useful formalism that not only helps in our understanding of the
underlying dynamics among various factors (e.g. \textit{the
influence of the completeness of the training dataset, the
complexity of processes etc.}), but also provides a practical
guideline as to how to develop efficient training procedures for
AID detectors to make training faster and how to identify the
intrusion context in the intrusive dataset to study intrusion
characteristics. Contradictory to our general concept that more
training audit trails will lead to more efficient stide detectors,
the experimental result show that there are critical sections in
the training audit trails, which are important to stide
efficiency. Our trimming scheme is to find such critical sections
in the training audit trails. Ultimately, the framework gives
guidelines for selecting stide for intrusion detection (i.e., what
are the applicable scenarios of stide). Though our discussion is
based on stide, the framework provides insights that are more
generally applicable to any sequence analysis based AID technique.

\textit{Related work.} Sequence time-delay embedding or
(\textit{stide}), was first proposed by Forrest et al
\cite{Forrest96SenseSelf} for privileged Unix processes. The
method is an instance of a computer immunology system to protect
the computer systems using the principles of natural immune
systems. However, throughout the series of papers by Forrest et al
\cite{Hofmeyr98SeqSystemCall} \cite{Stillerman99DistributedAppID}
\cite{Warrender99Comparison}, there is this ``\textit{magic number
6}" \cite{Marceau00MultipleLengthNgrams} which is empirically
determined to be the length of stide detectors to obtain effective
detection of all anomalies in intrusive datasets. This so called
`\textbf{Why six?}' problem \cite{Tan01Why6} for \emph{stide} has
stimulated a lot of research \cite{Lee01InfoTheoreticAD}
\cite{Kymie03DetermineOpLmts}. Finally, Tan et al.
\cite{Kymie03DetermineOpLmts} \cite{Tan01Why6} showed that the
correct answer to the problem lies in the fact that the lower
bound on the stide detector length is determined by the length of
the minimum foreign sequence(s) in the intrusive dataset.

However, Tan et al. \cite{Kymie03DetermineOpLmts} fell short of
providing a comprehensive framework that systematically analyzes
the interactions among various factors affecting the operational
limits of stide detectors. In particular, it fails to establish
the effects of incompleteness of the training dataset on the
effectiveness of stide. Though it is generally understood that
more completeness of the training audit trails leads to higher
detection rate and lower false alarm rate, a quantitative
relationship among them is not available in the literatures. In
practice, even though there exist some techniques to generate
near-complete training dataset
\cite{Debar98RefAuditInfoGeneration}, it is worth studying the
effect of the completeness of the training dataset on the
performance of stide detectors for the following reasons:
\begin{enumerate}
\item It is difficult to collect a complete training dataset even
with the completeness-guarantee techniques;

\item The existing completeness-guarantee techniques are only
specific for the system-call-based events in a host. However, as a
general technique, stide is not only applicable to system call
sequences in a host, but also other event sequences in diverse
environments, such as networks;

\item In a converse manner, the knowledge of the precise influence
of the completeness of the training dataset on the efficiency of
stide detectors will be useful to propose or improve
completeness-guarantee techniques;

\item The`\textit{concept drifting}' problem in the normal model
for anomaly-based intrusion detection tends to make the normal
model always incomplete;

\item Most important of all, the tradeoff between the completeness
of the training dataset and the efficiency of the stide detector
needs to be quantified. It is a common sense that, as the training
dataset approaches more completeness, more efforts are needed to
achieve any information gain. Furthermore, it is possible that the
efficiency loss due to the incompleteness will be made up by
modeling generalization.
\end{enumerate}

\textit{Contributions of this paper.} The main contributions are
summarized below:
\begin{itemize}
\item A formal framework is proposed to determine the operational
limits for stide-like AID detectors. Under the framework, the
other techniques related to stide, namely, mimicry attacks,
information hiding techniques, variable length patterns, t-stide
and locality frame, are interpreted in a logical way.

\item Under the framework, a comprehensive solution to the `Why
six?' problem is achieved, which extends the one presented by Tan
et al. \cite{Tan01Why6}\cite{Kymie03DetermineOpLmts}.

\item The influence of the completeness of the training dataset on
stide efficiency is evaluated.

\item A methodology is derived from the formal framework for
trimming the training data for a specific detection performance by
identifying and eliminating non-critical sections since they yield
no additional information gain. This saves both training time and
space for storing training data.

\item A scheme for identifying the intrusion context is proposed,
and several useful findings from the minimum foreign sequences in
the intrusive dataset are reported, at least for AID techniques
based on sequence analysis.
\end{itemize}

The remaining paper is organized as follows. Section 2 gives the
notations and definitions to help the readers in understanding the
rest of the paper. In section 3, stide is briefly introduced and
expressed formally. The performance measures such as the
effectiveness, completeness and efficiency of an anomaly-based
intrusion detector are defined and several theorems on them are
presented and proved in section 4. In addition, the operational
limits for stide detectors are determined. In section 5, the
influence of the completeness of training dataset on stide
efficiency is evaluated, and then the intrusion context
identification scheme is proposed and evaluated using a typical
dataset. In the last section, conclusions are drawn and future
work on our framework is discussed.

\section{Notations and Definitions}
\subsection{Notations}
\textit{Sequences and Sequence Sets:} Let $\Sigma$ denote the
dataset for a process, which consists of event logs with the
identity of the associated running process. A sequence $S$ in
$\Sigma$ is an event series constituted by contiguous events in
$\Sigma$ with the same process identity, and its length is denoted
as $|S|$. Specially, $\phi$ is a sequence with length 0, and
$\Sigma$ itself is a sequence as well. $SS(\Sigma, l)$ denotes the
set of all the sequences of length $l$ ($l\geqslant 0$), which are
collected from $\Sigma$. Thus, $SS(\Sigma, 0)=\{\phi\}$.
Furthermore, $SS(\Sigma)=\bigcup_{l=0}^{+\infty}SS(\Sigma, l)$. In
any subset of $SS'(\Sigma)\subset SS(\Sigma)$,
$|SS'|_{min}(\Sigma)$\footnote{We use the notation $|...|$ to
represent the length of any member sequence in a sequence set,
instead of its size.} is the minimum length of all sequences in
$SS'(\Sigma)$, and $SS'_{min}(\Sigma)$ consists of the sequences
with length $|SS'|_{min}(\Sigma)$ in $SS'(\Sigma)$. As a special
case, $|SS'|_{min}(\Sigma)=0$ if $\phi\in SS'_{min}(\Sigma)$, and
$|SS|_{min}(\Sigma)=1$.
\begin{example}
Suppose $\Sigma=abc$. $ab$ is a sequence in $\Sigma$ with length
$2$, $SS(\Sigma, 2)=\{ab,bc\}$, and
$SS(\Sigma)=\{\phi,a,b,c,ab,bc,\\abc\}$. For a subset
$SS'(\Sigma)=\{b,c,ab,abc\}$, $|SS'|_{min}(\Sigma)=1$ and
$SS'_{min}(\Sigma)=\{b,c\}$
\end{example}

\textit{Set Operations:} For given datasets $\Sigma_{1}$ and
$\Sigma_{2}$ of a process and corresponding sequence sets
$SS(\Sigma_{1}, l)$ and $SS(\Sigma_{2}, l)$, the set operations
($\cup,\cap,-$) are defined as follows ($l\geqslant 0$):
\begin{eqnarray*}
&(1)&SS(\Sigma_{1}, l)\cup SS(\Sigma_{2}, l) \\&&=\{S|(S\in
SS(\Sigma_{1}, l)) \vee (S\in SS(\Sigma_{2}, l))\}\\
&(2)&SS(\Sigma_{1}, l)\cap SS(\Sigma_{2}, l) \\&&=\{S|(S\in
SS(\Sigma_{1}, l)) \wedge (S\in
SS(\Sigma_{2}, l))\}\\
&(3)&SS(\Sigma_{1}, l)-SS(\Sigma_{2}, l) \\&&=\{S|(S\in
SS(\Sigma_{1}, l)) \wedge (S\not\in SS(\Sigma_{2}, l))\}
\end{eqnarray*}
In addition, $\Sigma_{1}\odot\Sigma_{2}$ is a special
concatenation of the datasets $\Sigma_{1}$ and $\Sigma_{2}$, such
that there is no sequence in $SS(\Sigma_{1}\odot\Sigma_{2},l)$, in
which some events belong to $\Sigma_{1}$ and other events belong
to $\Sigma_{1}$. This is because the process identity in
$\Sigma_{1}$ is different from that in $\Sigma_{2}$. Therefore,
$SS(\Sigma_{1}\odot\Sigma_{2},l)=SS(\Sigma_{1},l)\cup
SS(\Sigma_{2},l)$.
\begin{example}
Suppose that $\Sigma_{1}=abc$ and $\Sigma_{2}=ab$.
$SS(\Sigma_{1},\\2)=\{ab,bc\}$, and $SS(\Sigma_{2},2) =\{ab\}$.
Thus, the set operations $SS(\Sigma_{1}, 2)\cup SS(\Sigma_{2},
2)=\{ab,bc\}$, $SS(\Sigma_{1}, 2)\cap SS(\Sigma_{2}, 2)=\{ab\}$,
and $SS(\Sigma_{1}, 2)-SS(\Sigma_{2}, 2)=\{bc\}$.
$\Sigma_{1}\odot\Sigma_{2}=abc;ab$.
\end{example}

\textit{Supersequence and Subsequence:} If $S_{sub}$ is a
contiguous subsequence of $S$ and $|S|-|S_{sub}|=k$, then
$S_{sub}$ is said to be a $k$-order subsequence of $S$, and
denoted as $S_{sub}\preccurlyeq_{k} S$. Similarly,
$S_{sup}\succcurlyeq_{k} S$ denotes that $S_{sup}$ is a $k$-order
supersequence in which $S$ is a contiguous subsequence, and
$|S_{sub}|-|S|=k$. It is worth noting that
$\phi\preccurlyeq_{|S|}S$, and $S\succcurlyeq_{|S|}\phi$. For
example, $ab\preccurlyeq_{1}abc$, $a\preccurlyeq_{2}abc$ and
$ab\succcurlyeq_{1}a$. In addition, the terms \textit{subsequence}
and \textit{supersequence} will always imply contiguity in this
paper, such that $ac\not\preccurlyeq_{1}abc$.

\subsection{Definitions}
Central to our framework are the twin concepts of the
\textit{minimum foreign sequence} -- \textbf{MFS}, and the
\textit{maximum self sequence} -- \textbf{MSS}. Their definitions,
expressions and relation are given below.
\subsubsection{Foreign sequences and self sequences}
Let  $\Sigma_{ref}$ be the \emph{reference dataset}, and
$\Sigma_{tgt}$ be the \emph{target dataset}. For any sequence
$S\in SS(\Sigma_{tgt})$, if $S$ is also in $SS(\Sigma_{ref})$, $S$
will be called a \textit{self sequence}, otherwise, it is a
\textit{foreign sequence} to $\Sigma_{ref}$. Furthermore,
$FRGN(\Sigma_{tgt}| \Sigma_{ref})$ is defined as the set of
foreign sequences of $\Sigma_{tgt}$ w.r.t. $\Sigma_{ref}$.
Similarly, the set of self sequences is defined as
$SELF(\Sigma_{tgt}|\\ \Sigma_{ref})$. Mathematically,
\begin{eqnarray}
FRGN(\Sigma_{tgt}|
\Sigma_{ref})=\mathop{\cup_{l=1}^{+\infty}}SS(\Sigma_{tgt},
l)-SS(\Sigma_{ref}, l)\\
SELF(\Sigma_{tgt}|
\Sigma_{ref})=\mathop{\cup_{l=1}^{+\infty}}SS(\Sigma_{tgt}, l)\cap
SS(\Sigma_{ref}, l)
\end{eqnarray}
Thus, \begin{math}FRGN(\Sigma_{tgt}| \Sigma_{ref})\cup
SELF(\Sigma_{tgt}| \Sigma_{ref})=SS(\Sigma_{tgt})\end{math}.

A sequence $S$ in $FRGN(\Sigma_{tgt}| \Sigma_{ref})$ will be
called a \textit{minimum foreign sequence (MFS)}
\cite{Kymie03DetermineOpLmts} if none of its subsequences is in
$FRGN(\Sigma_{tgt}| \Sigma_{ref})$, i.e. all of its subsequences
are in $SELF(\Sigma_{tgt}|\Sigma_{ref})$. The set of all minimum
foreign sequences is denoted as $MFS(\Sigma_{tgt}| \Sigma_{ref})$.
On the other hand, for any sequence $S$ in $SELF(\Sigma_{tgt}|
\Sigma_{ref})$, if there exists one $1$-order supersequence that
is not included in $SELF(\Sigma_{tgt}| \Sigma_{ref})$ (i.e., it is
in $FRGN(\Sigma_{tgt}| \Sigma_{ref})$), then it will be called a
\textit{maximum self sequence (MSS)}. The set of all maximum self
sequences is denoted as $MSS(\Sigma_{tgt}| \Sigma_{ref})$.
Formally, they can be expressed as:
\begin{eqnarray}
\label{eqn:MFS-MSS}
&&\hspace{-20pt}MFS(\Sigma_{tgt}|\Sigma_{ref})\nonumber\\
&=&\hspace{-10pt}\{S|\forall S(S\in
FRGN(\Sigma_{tgt}|\Sigma_{ref}))\wedge(\forall S'\forall k(S'\in
SS(\Sigma_{tgt}))\nonumber\\&&\wedge(S'\preccurlyeq_{k}
S\rightarrow S'\not\in
FRGN(\Sigma_{tgt}| \Sigma_{ref})))\}\\
&&\hspace{-20pt}MSS(\Sigma_{tgt}| \Sigma_{ref})\nonumber\\
&=&\hspace{-10pt}\{S|\forall S(S\in SELF(\Sigma_{tgt}|
\Sigma_{ref}))\wedge(\exists S'(S'\in
SS(\Sigma_{tgt}))\nonumber\\&&\wedge (S'\succcurlyeq_{1} S)\wedge(
S'\not\in SELF(\Sigma_{tgt}| \Sigma_{ref})))\}
\end{eqnarray}
From these definitions, $MFS(\Sigma_{tgt}|\Sigma_{ref})\subset
SS(\Sigma_{tgt})$ and $MSS(\Sigma_{tgt}| \Sigma_{ref})\subset
SS(\Sigma_{tgt})$. Furthermore, based on above notations,
$|MFS|_{min}(\Sigma_{tgt}| \Sigma_{ref})\geqslant 1$,
$|MSS|_{min}(\Sigma_{tgt}| \Sigma_{ref})\geqslant 0$. Specially,
if $MFS(\Sigma_{tgt}|\Sigma_{ref})=\Phi$,
$|MFS|_{min}(\Sigma_{tgt}| \Sigma_{ref})=+\infty$. The same
property can be applied to $MSS(\Sigma_{tgt}|\Sigma_{ref})$.

\begin{example}
\label{epl:FRGNSELF} Suppose that $\Sigma_{ref}=abc$,
$\Sigma_{tgt}=abaa$. The sequence sets of these two datasets are:
$SS(\Sigma_{ref})=\{\phi,a,b,c,\\ab,bc,abc\}$,
$SS(\Sigma_{tgt})=\{\phi,a,b,ab,ba,aa,aba,baa,abaa\}$. Next,
\begin{eqnarray*}
FRGN(\Sigma_{tgt}|
\Sigma_{ref})&=&\{ba,aa,aba,baa,abaa\}\\
SELF(\Sigma_{tgt}| \Sigma_{ref})&=&\{\phi,a,b,ab\}
\end{eqnarray*}
Finally, we can deduce:
\begin{eqnarray*}
MFS(\Sigma_{tgt}|\Sigma_{ref})&=&\{ba,aa\}\\
MSS(\Sigma_{tgt}|\Sigma_{ref})&=&\{a,b,ab\}\\
MFS_{min}(\Sigma_{tgt}|\Sigma_{ref})&=&\{ba,aa\}\\
MSS_{min}(\Sigma_{tgt}|\Sigma_{ref})&=&\{a,b\}\\
|MFS|_{min}(\Sigma_{tgt}|\Sigma_{ref})&=&2\\
|MSS|_{min}(\Sigma_{tgt}|\Sigma_{ref})&=&1
\end{eqnarray*}
\end{example}

\subsubsection{Relation between MFS and MSS}
One relationship between MFSs and MSSs of two datasets
$\Sigma_{ref}$ and $\Sigma_{tgt}$ is given by the following
theorem\footnote{To save space, all proofs of the theorems in this
paper are provided in our (extended) technical report
\cite{Li04stideFramework} at
http://www.cais.ntu.edu.sg/home/technical\_reports\_2004.jsp.}.
\begin{theorem}
\label{theorem:MFS=MSS+1} For two datasets $\Sigma_{ref}$ and
$\Sigma_{tgt}$ of a process, the following relation holds.
\begin{eqnarray}
|MSS|_{min}(\Sigma_{tgt}| \Sigma_{ref})=|MFS|_{min}(\Sigma_{tgt}|
\Sigma_{ref})-1
\end{eqnarray}
\end{theorem}
\begin{example}
In Example~\ref{epl:FRGNSELF}, it is obvious that\\
$|MSS|_{min}(\Sigma_{tgt}|\Sigma_{ref})=|MFS|_{min}(\Sigma_{tgt}|\Sigma_{ref})-1=1$.
\end{example}

\section{A formal description of stide}
In the experimental setup for stide \cite{Forrest96SenseSelf}
\cite{Warrender99Comparison}, there are two datasets for every
process, the normal dataset $\Sigma_{nml}$, and the intrusive
dataset $\Sigma_{int}$, which are defined below.
\begin{definition}[Normal Dataset]
The normal dataset is a dataset $\Sigma_{nml}$ that is utilized to
train the normal model of a process for stide, and it {\bfseries
MUST} be collected in the normal run of the process without any
intrusion.
\end{definition}
\begin{definition}[Intrusive Dataset]
\label{def:intruDS} The intrusive dataset $\Sigma_{int}$ is a
dataset that is collected when one or more intrusions were
occurring during the runs of a process.
\end{definition}

In terms of these two datasets from the same process, stide can be
formally described as follows. Let $\omega(\geqslant 1)$ denote
the size of the detector window. In the modeling phase, the normal
model of the process is obtained as: $SS(\Sigma_{nml}, \omega)$.
Then, in the detecting phase, the foreign sequences in the
intrusive dataset $\Sigma_{int}$, $FS(\Sigma_{int}| \Sigma_{nml},
\omega)$, are enumerated:
\begin{eqnarray*}
FS(\Sigma_{int}|\Sigma_{nml}, \omega)=SS(\Sigma_{int},
\omega)-SS(\Sigma_{nml}, \omega)
\end{eqnarray*}
If $FS(\Sigma_{int}| \Sigma_{nml}, \omega)\neq\Phi$, the
intrusion(s) in the intrusive dataset $\Sigma_{int}$ can be
detected with the detector length $\omega$
\cite{Hofmeyr98SeqSystemCall} \cite{Hofmeyr00ArchitectureAIS}
\cite{Marceau00MultipleLengthNgrams} \cite{Kymie03DetermineOpLmts}
\cite{Warrender99Comparison}\footnote{It is notable that most of
these research studies only apply stide to system-call based
sequences in a host as does the original proposal for
stide\cite{Hofmeyr98SeqSystemCall}. However, in principle, stide
is applicable to other environments as well. Therefore, in our
formal framework, it will not be specific for any environment,
which is also one of our objectives to formalize the stide
technique.}. It is evident that the sequence set $FS(\Sigma_{int}|
\Sigma_{nml}, \omega)$ is strongly related to
$MFS(\Sigma_{int}|\Sigma_{nml})$ via
$|MFS|_{min}(\Sigma_{int}|\Sigma_{nml})$:
\begin{eqnarray}
|MFS|_{min}(\Sigma_{int}| \Sigma_{nml})\leqslant
\omega\Leftrightarrow FS(\Sigma_{int}| \Sigma_{nml}, \omega)\neq
\Phi
\end{eqnarray}
In its formal proposal \cite{Hofmeyr98SeqSystemCall}, a Locality
Frame Count (LFC) function is applied to smooth the noise, or to
filter the false alarms in the process by summing up the number of
foreign sequences found within the span of a locality frame.
However, the LFC function does not add to or compensate for the
detecting ability, or failure/shortcoming of the stide detector,
and it will not be used in our framework. As an application of our
framework, it will be interpreted later.

Practically, even though the underlying principle is very simple,
stide can detect most of the intrusions into the processes
(\emph{the datasets from UNM} \cite{Dataset94UNM}). For this
reason, it is accepted as a typical and effective anomaly-based
intrusion detector in many research studies.

\section{A formal framework for stide}
In general, the efficiency of an intrusion detection technique is
determined by both false positives
\cite{Axelsson99BaseRateFallacy} and false negatives. The
incompleteness of the normal model is well-known as the main cause
for the false positives in an AID detector \cite{Li04ACI}
\cite{Axelsson99BaseRateFallacy}. However, in most of the research
studies on stide \cite{Forrest96SenseSelf}
\cite{Kymie03DetermineOpLmts}, the completeness is not adequately
considered when evaluating the efficiency of stide detectors. In
other words, there is an implicit assumption that the normal
dataset is complete in the sense that it includes all the normal
behaviors of a process. As a result, the issue of false positives
has been completely ignored. However, as indicated in the first
section, such completeness of the normal dataset is difficult to
verify, and there is no effective method to guarantee it.

In our framework, the implicit assumption about the completeness
of the normal dataset is discarded, and the normal dataset is
regarded as the training dataset $\Sigma_{trn}$ to build the known
normal model of the resource. At the same time, a test dataset
$\Sigma_{tst}$ is introduced to evaluate the completeness of the
training dataset $\Sigma_{trn}$. The function of the test dataset
is to evaluate the ability of the detector to correctly identify
normal data as such without generating  false positives. Thus, the
test dataset must be collected during a normal run of a process
without any intrusion as well. To some extent, our methodology
corresponds to the actual scenarios where it is difficult to
collect all the normal behaviors of a computing resource, and
there are always false positives when the normal behaviors of the
process are examined by an AID detector. In addition, without loss
of generality, we assume that the audit trails in $\Sigma_{int}$
is caused by only one intrusion.

\subsection{A critical look at stide performance}
In our formal framework for stide, with the detector window size
$\omega$, the normal model $SS(\Sigma_{trn}, \omega)$ is first
gleaned from $\Sigma_{trn}$. Based on its detection results on
$\Sigma_{tst}$ and $\Sigma_{int}$, all the sequences are
classified as follows. The outcome of a detection process can be
divided into four categories depending on the true nature of the
data and the correctness of the detection result. These are shown
in Table~\ref{tbl:4Detection}.
\begin{table}[h]
\centering
  \caption{Four detection scenarios.}\label{tbl:4Detection}
  \begin{scriptsize}
  \begin{tabular}{|c||c|c|}
    \hline                          & \textsc{intrusive sequence} & \textsc{normal sequence} \\
    \hline\hline \textsc{alarm}     & True Positive      & False Positive \\
    \hline       \textsc{non-alarm} & False Negative     & True Negative \\
    \hline
  \end{tabular}
  \end{scriptsize}
\end{table}
Therefore, according to whether a sequence matches the normal
model $SS(\Sigma_{trn}, \omega)$, $SS(\Sigma_{tst},\omega)$ can be
split into two subsets: False Positive Sequence Set
(\textit{denoted as} $FPSS(\Sigma_{tst}|\Sigma_{trn},\omega)$),
and True Negative Sequence Set (\textit{denoted as}
$TNSS(\Sigma_{tst}| \Sigma_{trn}, \omega)$). Similarly, depending
on the detection outcome, the intrusive sequence set
$SS(\Sigma_{int}, \omega)$ can be split into two subsets: False
Negative Sequence Set (\textit{denoted as} $FNSS(\Sigma_{int}|
\Sigma_{trn}, \omega)$), and True Positive Sequence Set
(\textit{denoted as} $TPSS(\Sigma_{int}| \Sigma_{trn},\\
\omega)$). Using our earlier notations, we can write the following
definitions of the above four sequence subsets:
\begin{eqnarray*}
FPSS(\Sigma_{tst}|\Sigma_{trn}, \omega)&=&SS(\Sigma_{tst},
\omega)-SS(\Sigma_{trn}, \omega)\\
TNSS(\Sigma_{tst}|\Sigma_{trn}, \omega)&=&SS(\Sigma_{tst},
\omega)\cap
SS(\Sigma_{trn}, \omega)\\
TPSS(\Sigma_{int}| \Sigma_{trn}, \omega)&=&SS(\Sigma_{int},
\omega)-
SS(\Sigma_{trn}, \omega)\\
FNSS(\Sigma_{int}| \Sigma_{trn}, \omega)&=&SS(\Sigma_{int},
\omega)\cap SS(\Sigma_{trn},\omega)
\end{eqnarray*}
Furthermore,
\begin{eqnarray*}
FRGN(\Sigma_{int}| \Sigma_{trn}) &=&
\mathop{\cup_{\omega=1}^{+\infty}}TPSS(\Sigma_{int}|
\Sigma_{trn}, \omega)\\
SELF(\Sigma_{int}| \Sigma_{trn}) &=&
\mathop{\cup_{\omega=1}^{+\infty}}FNSS(\Sigma_{int}| \Sigma_{trn}, \omega)\\
FRGN(\Sigma_{tst}| \Sigma_{trn}) &=&
\mathop{\cup_{\omega=1}^{+\infty}}FPSS(\Sigma_{tst}|
\Sigma_{trn}, \omega)\\
SELF(\Sigma_{tst}| \Sigma_{trn}) &=&
\mathop{\cup_{\omega=1}^{+\infty}}TNSS(\Sigma_{tst}| \Sigma_{trn},
\omega)
\end{eqnarray*}

Next, according to the sequences in these four categories, we will
define two aspects of stide performance, namely effectiveness and
completeness. Finally, we will give the definitions and conditions
for an efficient stide detector.

\subsubsection{Effectiveness of a stide detector}
\begin{definition}[Effectiveness]
A stide detector with detector window $\omega$ is
\textbf{effective} to detect the intrusion in $\Sigma_{int}$ if
there is at least one sequence in the intrusive sequence set
$SS(\Sigma_{int}, \omega)$, which is detected as a true positive,
i.e., $TPSS(\Sigma_{int}| \Sigma_{trn}, \omega)\neq\Phi$.
\end{definition}

To detect an intrusion effectively, the relation between stide
detector window size $\omega$ and the intrusion characteristics is
critical to choose a proper $\omega$ for stide, and it is stated
in the following theorem.
\begin{theorem}\label{theorem:effective}
Let us assume that there are a training dataset $\Sigma_{trn}$ and
an intrusive dataset $\Sigma_{int}$ of a process. A stide detector
of length $\omega$, built from $\Sigma_{trn}$, is effective w.r.t.
$\Sigma_{int}$, iff
\begin{equation}
\omega\geqslant |MFS|_{min}(\Sigma_{int} |  \Sigma_{trn})
\end{equation}
\end{theorem}
\begin{example}
Suppose that $\Sigma_{trn}=aba$, and $\Sigma_{int}=ababa$. Then,
$MFS_{min}(\Sigma_{int}|\Sigma_{trn})=\{bab\}$, and
$|MFS|_{min}(\Sigma_{int}|\Sigma_{trn})\\=3$. As
$SS(\Sigma_{int},1)=SS(\Sigma_{trn},1)=\{a,b\}$, and
$SS(\Sigma_{int},2)=SS(\Sigma_{trn},2)=\{ab,ba\}$,
$TPSS(\Sigma_{int}|\Sigma_{trn}, 1)=\Phi$ and
$TPSS\\(\Sigma_{int}|\Sigma_{trn}, 2)=\Phi$. But
$TPSS(\Sigma_{int}|\Sigma_{trn}, 3)=\{bab\}\neq\Phi$ Thus, only if
$\omega\geq 3$, the intrusion in $\Sigma_{int}$ will be detected
by stide effectively.
\end{example}
Note that Theorem~\ref{theorem:effective} merely summarizes the
conclusion of Tan et al. \cite{Kymie03DetermineOpLmts}, but in our
framework, it is rather straightforward to prove its validity.

\subsubsection{Completeness of a stide detector}
\begin{definition}[Completeness]
A stide detector with detector window $\omega$ is
\textbf{complete} if the underlying normal model built from a
training dataset $\Sigma_{trn}$ is complete. In other words, the
sequence subsets $TNSS(\Sigma_{tst}|\Sigma_{trn},
\omega)=SS(\Sigma_{tst}, \omega)$, and thus
$FPSS(\Sigma_{tst}|\Sigma_{trn}, \omega)=\Phi$.
\end{definition}

Due to the base-rate fallacy \cite{Axelsson99BaseRateFallacy}, the
completeness of a stide detector is also critical for its
application. The following theorem establishes the conditions for
the completeness of a stide detector in terms of the detector
window size $\omega$.
\begin{theorem}
\label{theorem:complete} Let us assume that there are a training
dataset $\Sigma_{trn}$ and a test dataset $\Sigma_{tst}$ of a
process. A stide detector of length $\omega$, built from
$\Sigma_{trn}$, is complete w.r.t. $\Sigma_{tst}$, iff
\begin{eqnarray}
\omega\leqslant |MSS|_{min}(\Sigma_{tst}| \Sigma_{trn})
\end{eqnarray}
\end{theorem}
\begin{example}
Suppose that $\Sigma_{trn}=aba$, and $\Sigma_{tst}=baba$. Then,
$MSS_{min}(\Sigma_{tst}|\Sigma_{trn})=\{ba,ab\}$, and
$|MSS|_{min}(\Sigma_{tst}|\\\Sigma_{trn})=2$. As
$SS(\Sigma_{tst},1)=SS(\Sigma_{trn},1)=\{a,b\}$, and
$SS(\Sigma_{tst},2)=SS(\Sigma_{trn},2)=\{ab,ba\}$,
$FPSS(\Sigma_{tst}|\Sigma_{trn}, 1)=\Phi$, and
$FPSS(\Sigma_{tst}|\Sigma_{trn}, 2)=\Phi$. On the other hand,
$FPSS\\(\Sigma_{tst}|\Sigma_{trn}, 3)=\{bab\}\neq\Phi$. Thus, only
if $\omega\leq 2$, the stide detector built from $\Sigma_{trn}$ is
complete w.r.t $\Sigma_{tst}$.
\end{example}

\begin{corollary}
\label{theorem:nocomplete} For a training dataset $\Sigma_{trn}$
and a test dataset $\Sigma_{tst}$ of a process, if
$|MSS|_{min}(\Sigma_{tst}|\Sigma_{trn})=0$, there are no complete
stide detectors built from $\Sigma_{trn}$ w.r.t. $\Sigma_{tst}$.
\end{corollary}

\subsubsection{Efficient stide detectors}
\begin{definition}[Efficiency]
A stide detector with detector window size $\omega$ is
\textbf{efficient} w.r.t $\Sigma_{tst}$ and $\Sigma_{int}$ if it
is effective to detect the intrusion in $\Sigma_{int}$, and it is
complete in detecting $\Sigma_{tst}$.
\end{definition}
It is easy to conclude that an efficient stide detector will not
produce any false positives when analyzing $\Sigma_{tst}$, and it
will produce true positives when analyzing $\Sigma_{int}$. We are
now in a position to state the condition for a stide detector to
be efficient, which is expressed by the following theorem.

\begin{theorem}
\label{theorem:MSS-MFSefficient} Given a training dataset
$\Sigma_{trn}$, a test dataset $\Sigma_{tst}$, and an intrusive
dataset $\Sigma_{int}$, a stide detector with the detection window
size $\omega$, obtained using $\Sigma_{trn}$, is efficient w.r.t.
$\Sigma_{tst}$ and $\Sigma_{int}$ iff
\begin{equation}
\label{eqn:EfficientIFF}
|MFS|_{min}(\Sigma_{int}|\Sigma_{trn})\leqslant \omega \leqslant
|MSS|_{min}(\Sigma_{tst}| \Sigma_{trn})
\end{equation}
\end{theorem}
\begin{proof}
It can be inferred from Theorem~\ref{theorem:effective} and
\ref{theorem:complete}.
\end{proof}
\begin{example}
Suppose that $\Sigma_{trn}=aba$, $\Sigma_{tst}=baba$, and
$\Sigma_{int}=abc$. Then,
$MSS_{min}(\Sigma_{tst}|\Sigma_{trn})=\{ba,ab\}$, and
$MFS(\Sigma_{int}|\Sigma_{trn})=\{c\}$, thus
$|MSS|_{min}(\Sigma_{tst}|\Sigma_{trn})=2$ and
$|MFS|_{min}(\Sigma_{int}|\Sigma_{trn})=1$. For a stide detector
with length $\omega$, to be complete, $\omega\leq 2$, and to be
effective, $\omega\geq 1$. Finally, we can get $1\leq\omega\leq
2$.
\end{example}
\begin{figure}[h]
\centering \epsfig{file=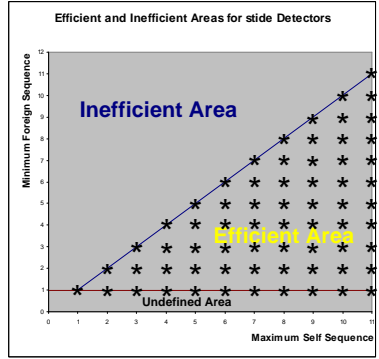,width=2.0in}
\caption{Efficient and inefficient areas for stide detectors on
the training dataset $\Sigma_{trn}$.} \label{fig:MSS-MFS}
\end{figure}
Figure~\ref{fig:MSS-MFS} shows the area determined by
$MFS_{min}(\Sigma_{int}|\Sigma_{trn})$ and
$MSS_{min}(\Sigma_{tst}|\Sigma_{trn})$ where efficient stide
detectors must belong. If the point determined by
$|MFS|_{min}(\Sigma_{int}|\Sigma_{trn})$ and
$|MSS|_{min}(\Sigma_{tst}|\Sigma_{trn})$ is in the efficient area,
it is possible to find one or more efficient stide detector(s).
Otherwise, no efficient stide detector can be found. Note that,
there is one undefined area since $|MFS|_{min}(\Sigma_{int} |
\Sigma_{trn})\geqslant 1$.

The following corollary, drawn from the above theorem, explicitly
defines the operational limits of a stide detector.
\begin{corollary}
\label{corol:Noefficient} For a training dataset $\Sigma_{trn}$, a
test dataset $\Sigma_{tst}$, and an intrusive dataset
$\Sigma_{int}$ of a process, the following hold:
\begin{enumerate}
\item [(a).]If $|MSS|_{min}(\Sigma_{tst}| \Sigma_{trn})<
|MFS|_{min}(\Sigma_{int} | \Sigma_{trn})$, there are no efficient
stide detectors w.r.t. $\Sigma_{tst}$ and $\Sigma_{int}$.

\item [(b).] With a detector window $\omega$, if $\omega\geqslant
|MFS|_{min}(\Sigma_{int} |  \Sigma_{trn})$, and $\omega\geqslant
|MSS|_{min}(\Sigma_{tst} |  \Sigma_{trn})$, the stide detector
built by $\Sigma_{trn}$ is effective, but not efficient w.r.t.
$\Sigma_{tst}$ and $\Sigma_{int}$.

\item [(c).] With a detector window $\omega$, if $\omega\leqslant
|MFS|_{min}(\Sigma_{int} |  \Sigma_{trn})$, and $\omega\leqslant
|MSS|_{min}(\Sigma_{tst} |  \Sigma_{trn})$, the stide detector
built by $\Sigma_{trn}$ is complete, but not efficient w.r.t.
$\Sigma_{tst}$ and $\Sigma_{int}$.
\end{enumerate}
\end{corollary}

\subsection{Completeness of the training dataset vs. stide efficiency}
From their definitions, $MFS(\Sigma_{int}|\Sigma_{trn})$ and
$MSS(\Sigma_{tst}|\Sigma_{trn})$ will be affected by the
completeness of the training dataset to a large extent. Therefore,
according to Theorem~\ref{theorem:MSS-MFSefficient}, the
completeness of the training dataset is critical to stide
efficiency.
\begin{figure}[h]
\centering \epsfig{file=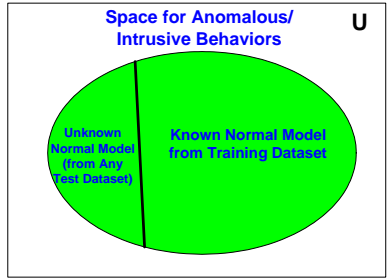,width=2.0in}
\caption{Behavior spaces for \textit{stide}: the known and unknown
normal behavior models, and the intrusive behavior space.}
\label{fig:ModelSpace}
\end{figure}
In Figure~\ref{fig:ModelSpace}, the universe of all possible
sequences is referred to as $U$, in which the known and unknown
normal models are only complementary parts of the complete normal
model. Outside the complete normal model is the intrusive behavior
space for the known and unknown intrusions. However, in stide, all
the sequences from the training dataset are regarded as normal (in
the known normal model), and other sequences lying outside the
training dataset are considered anomalous. Obviously, the unknown
normal model is critical for its efficiency. Thus, in our
following analysis, we assume that the test dataset $\Sigma_{tst}$
incorporates the whole unknown normal model
(Figure~\ref{fig:ModelSpace}). Even though the assumption can not
be achieved in the real deployment, it is reasonable in analyzing
stide efficiency. Given this framework, let's examine the scenario
in which stide is suitable for detecting the intrusions into a
resource.

\subsubsection{MSSs in the test dataset}
Based on Equation~\ref{eqn:MFS-MSS},
$MSS(\Sigma_{tst}|\Sigma_{trn})$ is deduced as:
\begin{eqnarray*}
\label{eqn:MSSIncomplete}
&&\hspace{-20pt}MSS(\Sigma_{tst}|\Sigma_{trn})\\&=&\{S|\forall
S(S\in SELF(\Sigma_{tst}| \Sigma_{trn}))\wedge(\exists S'(S'\in
SS(\Sigma_{tst}))\nonumber\\&&\wedge (S'\succcurlyeq_{1} S)\wedge(
S'\not\in SELF(\Sigma_{tst}| \Sigma_{trn})))\}\\
&=&\{S|\forall S(S\in SELF(\Sigma_{tst}|
\Sigma_{trn}))\wedge(\exists S'(S'\in
SS(\Sigma_{tst}))\nonumber\\&&\wedge (S'\succcurlyeq_{1} S)\wedge(
S'\in FRGN(\Sigma_{tst}| \Sigma_{trn})))\}\nonumber
\end{eqnarray*}
Thus, $MSS(\Sigma_{tst} | \Sigma_{trn})$ is affected by
$FRGN(\Sigma_{tst} | \Sigma_{trn})$, which is in the unknown
normal model (Figure~\ref{fig:ModelSpace}). Theoretically, for any
$\Sigma_{trn}$ and $\Sigma_{tst}$, if $\Sigma_{trn}$ is not
complete, $|MSS|_{min}(\Sigma_{tst}|\\\Sigma_{trn})$ can vary from
0 to $+\infty$ (\textbf{Note}: $+\infty$ here indicates a
potentially large number bounded by the length of the dataset
$\Sigma_{tst}$).

\subsubsection{MFSs in the intrusive dataset}
Let's first define one more concept in our framework:
\begin{definition} \label{def:MFPS} The common false
positive sequence set in the intrusive dataset
$CFPS(\Sigma_{int},\Sigma_{tst}| \Sigma_{trn})$ is:
\begin{eqnarray*}
\label{eqn:MFPS}
CFPS(\Sigma_{int},\Sigma_{tst}| \Sigma_{trn})=
FRGN(\Sigma_{tst}|\Sigma_{trn})\cap SS(\Sigma_{int})
\end{eqnarray*}
\end{definition}
It is obvious that $CFPS(\Sigma_{int},\Sigma_{tst}|
\Sigma_{trn})\subset SS(\Sigma_{int})$.

\begin{example}
\label{epl:CFPS} Suppose that $\Sigma_{trn}=ljk$,
$\Sigma_{tst}=jkl$ and $\Sigma_{int}=ckl$. Based on sequence set
definition, $SS(\Sigma_{trn})=\{\phi,l,j,k,lj,jk,ljk\}$,
$SS(\Sigma_{tst})=\{\phi,j,k,l,jk,kl,jkl\}$, and
$SS(\Sigma_{int})=\{\phi,c,k,l,ck,kl,ckl\}$. Next, we get
$FRGN(\Sigma_{tst}|\\\Sigma_{trn})=\{kl,jkl\}$. Therefore,
\begin{eqnarray*}
CFPS(\Sigma_{int},\Sigma_{tst}|
\Sigma_{trn})&=&\{kl\}\\
|CFPS|_{min}(\Sigma_{int},\Sigma_{tst}| \Sigma_{trn})&=&2
\end{eqnarray*}
\end{example}

The following theorem is deduced to determine what affects
$|MFS|_{min}(\Sigma_{int}|\Sigma_{trn})$ in our framework.
\begin{theorem}
\label{theorem:minMFPS-MFS} For the datasets $\Sigma_{trn}$,
$\Sigma_{tst}$, and $\Sigma_{int}$,
\begin{eqnarray}
\label{eqn:IncompleteMFS} &&\hspace{-20pt}|MFS|_{min}(\Sigma_{int}
| \Sigma_{trn})\nonumber\\ &=&
min(|CFPS|_{min}(\Sigma_{int},\Sigma_{tst}|
\Sigma_{trn}),\nonumber\\&&\hspace{20pt}|MFS|_{min}(\Sigma_{int}|\Sigma_{trn}\odot\Sigma_{tst}))
\end{eqnarray}
\end{theorem}
\begin{example}
\label{epl:MFS-CFPS} Take the same scenario in
example~\ref{epl:CFPS}. In it,
$MFS_{min}(\Sigma_{int}|\Sigma_{trn}\odot\Sigma_{tst})=\{c\}$, and
$|MFS|_{min}(\Sigma_{int}|\Sigma_{trn}\odot\Sigma_{tst})=1$. Thus,
we can determine that
$|MFS|_{min}(\Sigma_{int}|\Sigma_{trn})\\=min\{2,1\}=1$, which is
correct as $MFS_{min}(\Sigma_{int}|\Sigma_{trn})=\{c,kl\}$. On the
other hand, if $\Sigma_{int}=jkl$,
$SS(\Sigma_{int})=SS(\Sigma_{tst})$. Thus,
$CFPS(\Sigma_{int},\Sigma_{tst}| \Sigma_{trn})=\{kl,jkl\}$,
$MFS_{min}\\(\Sigma_{int}|\Sigma_{trn}\odot\Sigma_{tst})=\Phi$. As
$MFS_{min}(\Sigma_{int}|\Sigma_{trn})=\{kl\}$,
$|MFS|_{min}(\Sigma_{int}|\Sigma_{trn})=2=min(2,+\infty)$.
\end{example}

As indicated by the intrusive space in
Figure~\ref{fig:ModelSpace}, $|MFS|_{min}$\\
$(\Sigma_{int}|\Sigma_{trn}\odot\Sigma_{tst})$ reflects the
intrusion characteristics in a specific intrusive dataset
$\Sigma_{int}$, which does not depend on the completeness of the
training dataset. Thus, without regard to the completeness of
$\Sigma_{trn}$, $MFS(\Sigma_{int}|\Sigma_{trn}\odot\Sigma_{tst})$
and $SS(\Sigma_{int})$ will always be stable. According to
Theorem~\ref{theorem:minMFPS-MFS}, if
$|CFPS|_{min}(\Sigma_{int},\Sigma_{tst}|\Sigma_{trn})<
|MFS|_{min}(\Sigma_{int}|\Sigma_{trn}\odot\Sigma_{tst})$,
$|MFS|_{min}(\Sigma_{int}|\Sigma_{trn})$ will be affected through
the set $FRGN$\\ $(\Sigma_{tst}|\Sigma_{trn})$ as the completeness
of the training dataset increases.

In summary, both $MFS(\Sigma_{int}|\Sigma_{trn})$ and
$MSS(\Sigma_{tst}|\Sigma_{trn})$ are affected by the completeness
of the training dataset $\Sigma_{trn}$, i.e.
$FRGN(\Sigma_{tst}|\Sigma_{trn})$.

\subsubsection{Enhancing efficiency of a stide detector}
\begin{theorem}
\label{theorem:Tendency} Assume that, for a process, the training
dataset from which the known model is built, is $\Sigma_{trn}$,
the test dataset from which the whole unknown normal model is
built, is $\Sigma_{tst}$, and the intrusive dataset is
$\Sigma_{int}$. Then, there exist one or more efficient stide
detectors iff
\begin{equation}
|MSS|_{min}(\Sigma_{tst}|\Sigma_{trn})\geqslant
|MFS|_{min}(\Sigma_{int}|\Sigma_{trn}\odot\Sigma_{tst})
\end{equation}
\end{theorem}
\begin{example}
Take the two scenarios in examples~\ref{epl:CFPS} and
\ref{epl:MFS-CFPS}. If $\Sigma_{int}=ckl$,
$MSS_{min}(\Sigma_{tst}|\Sigma_{trn})=\{l,k\}$,
$MFS_{min}(\Sigma_{int}|\\\Sigma_{trn})=\{c\}$, and
$MFS_{min}(\Sigma_{int}|\Sigma_{trn}\odot\Sigma_{tst})=\{c\}$.
Thus, there exists only one efficient stide detector with length
$\omega=1$. However, if $\Sigma_{int}=jkl$,
$MSS_{min}(\Sigma_{tst}|\Sigma_{trn})=\{l,k\}$,
$MFS_{min}(\Sigma_{int}|\Sigma_{trn})=\{kl\}$, and thus, there do
not exist efficient stide detectors. At the same time, as
$MFS_{min}(\Sigma_{int}|\Sigma_{trn}\odot\Sigma_{tst})=\Phi$, and
$|MFS|_{min}(\Sigma_{int}|\Sigma_{trn}\odot\Sigma_{tst})=+\infty$,
the above equation in Theorem~\ref{theorem:Tendency} does not
hold.
\end{example}

What the above theorem tells us is that with increasing
completeness of the training dataset, the intrusion
characteristics reflected by $|MFS|_{min}(\Sigma_{int} |
\Sigma_{trn}\odot\Sigma_{tst})$ acts more and more as a threshold
for the efficiency of a stide detector. Therefore, sooner or
later, the intrusion must manifest itself in the intrusive dataset
with a finite (reasonably small) length of the MFSs, otherwise,
there will be no efficient stide detector for the intrusion.

The following corollary, which follows from the theorems
\ref{theorem:Tendency} and \ref{theorem:MSS-MFSefficient},
emphasizes the condition to build efficient stide detectors from a
training dataset:

\begin{corollary}
\label{corol:NoEffstide} Assume that, for a process, the training
dataset from which the known normal model is built, is
$\Sigma_{trn}$, the test dataset from which the unknown normal
model is constructed, is $\Sigma_{tst}$, and the intrusive dataset
is $\Sigma_{int}$. Then, if $|MFS|_{min}(\Sigma_{int} |
\Sigma_{trn})<
|MFS|_{min}(\Sigma_{int}|\Sigma_{trn}\odot\Sigma_{tst})$, there
are no efficient stide detectors.
\end{corollary}

Ideally, if the training dataset $\Sigma_{trn}$ is complete so
that it includes all the normal behaviors of a process (i.e., for
any $\Sigma_{tst}$,
$|MSS|_{min}(\Sigma_{tst}|\Sigma_{trn})=+\infty$),
$|CFPS|_{min}(\Sigma_{int}, \Sigma_{tst} |$\\
$\Sigma_{trn})=+\infty$. At the same time, the MFSs in an
intrusive dataset $\Sigma_{int}$, is in fact $MFS(\Sigma_{int} |
\Sigma_{trn}\odot\Sigma_{tst})$, which is the absolutely ideal
scenario. Definitely, under the ideal scenario, there will be
efficient stide detectors trained by the dataset $\Sigma_{trn}$.

\subsection{Interpretation of related work on stide}
Following the publication of stide, several research studies have
been published with criticisms and suggestions of improvement of
stide \cite{Wagner02MimicryAttacks} \cite{Tan02HidingIntrusion}
\cite{Warrender99Comparison} \cite{Wespi00VarGram}. Under our
proposed framework, they can be interpreted in a logical way to
determine their basic foundations.

\subsubsection{Mimicry attacks and intrusion information hiding}
The mimicry attacks are proposed by Wagner to show the weakness of
the stide technique \cite{Wagner02MimicryAttacks}. In a nutshell,
the proposed strategies to do mimicry attacks are:
\textit{intrusive behavior avoidance; waiting for the intrusive
behaviors accepted by normal model passively and actively;
replacing the system call parameters; inserting no-effect system
calls; creating equivalent variations of a given malicious
sequence}. Almost with the same principles, the information hiding
paradigm is also applied to indicate that the stide technique is
easy to be evaded \cite{Tan02HidingIntrusion}.

Utilizing our proposed framework, it is very obvious that all
these evading strategies are to make the minimum foreign sequences
of an intrusion as large as possible so that eventually it grows
beyond the length of stide detector window, making the stide
detector ineffective. From the viewpoint of information theory,
the information gain in the intrusive dataset (which is
manipulated by mimicry attacks or information hiding techniques)
is too small to be detected. Furthermore, since these two
techniques focus on applying stide to system call sequences in the
host-based systems with more or less strong assumptions, the only
conclusion that can be made is that the stide is not suitable for
detecting the mimicry attacks based on the system call sequences.
Therefore, at this point, no conclusion can be drawn about the
influence of these techniques on the efficiency of a general stide
detector (with a `\textit{good}' encoding replacing
`\textit{system call}'). Furthermore, from our following
experimental results, the large quantity of the minimum foreign
sequences discovered in every intrusive dataset will, to a large
extent, discourage mimicry attacks and intrusion information
hiding techniques,  because all of the minimum foreign sequences
must be mimicked and hidden to evade detection.

In addition, we found that it is possible to make stide
inefficient by getting control of one application and then nudging
it to generate smaller \textit{minimum common false positives}
during the run of an intrusion
(Theorem~\ref{theorem:minMFPS-MFS}). If the quantity of the false
positives are large enough during the intrusion, the stide
detector will be useless in detecting the intrusion due to the
base-rate fallacy\cite{Axelsson99BaseRateFallacy}.

\subsubsection{t-stide and variable length patterns}
As variations of stide, t-stide and variable-length patterns are
proposed in \cite{Warrender99Comparison} and
\cite{Wespi00VarGram}, and both of them utilize the frequency
information of each sequence. t-stide is very similar to stide
except that it discards infrequent sequences whose frequency is
smaller than a threshold $t$ \cite{Warrender99Comparison}. The
performance of t-stide is found to be unsatisfactory by the
author. Using our framework, the obvious reason for t-stide's
failure is that the discarded sequences will increase the
incompleteness of the training dataset, that will decrease its
detection efficiency. Furthermore, with the method of minimum
foreign sequence discovery discussed later in section
\ref{Application2}, the negative conclusion regarding t-stide can
be further explained by comparing the MFSs in the intrusive
dataset for stide and t-stide.

Since the principles for stide and variable-length patterns
\cite{Wespi00VarGram} are different, their comparisons will be
based on the detection performance by considering the fact that
only the patterns (or sequences) are used in the detection phase.
As indicated in our framework, the minimum foreign sequence is the
main characteristic left in the audit trails for the
sequence-related AID techniques. In the principles of
variable-length patterns, we noticed that if the minimum length of
the MFSs of an intrusion is larger than 1, the intrusion will be
easily ignored by variable-length patterns. For example, suppose
that the normal model for the variable-length method is \{ABCD,
CAE, FBD\}. If the minimum foreign sequence of an intrusion is
`DC', and the intrusive audit trail is `ABCDCAEFBD', the intrusion
will not be detected. As the experimental results in
\cite{Wespi00VarGram} are very good, we suspect that the MFS of
the chosen intrusions is 1, just like the `misconfiguration' for
`wu-ftpd' in our experiments described later.

\subsubsection{The significance of locality frame count}
For stide, following \cite{Warrender99Comparison}
\cite{Hofmeyr98SeqSystemCall}, the anomaly value of a trace is
derived from the number of mismatches occurring in a temporally
local region, called a locality frame (LF). Then, a locality frame
count (LFC) is used as a threshold to determine whether the
locality frame is anomalous in the trace.

Let us assume that the stide detector window is of length
$\omega$. Suppose that, in a locality frame of a trace, there are
at least $n$ MFSs: $MFS_{1}, MFS_{2}, \dots, MFS_{n}$ with lengths
smaller than $\omega$, and $MFS_{k}$ has the minimum length
$l_{k}$ among them. Since by definition, one MFS can not
completely include another MFS, the minimum number of mismatches
will take place when the MFS's are maximally overlapped, i.e.,
$MFS_{2}$ starts one event later than $MFS_{1}$, $MFS_{3}$ starts
one event later than $MFS_{2}$ and so on. In this pathological
case, the minimum number of anomalies detected in the LF should be
$\omega-l_{k}+n$. Therefore, for successful detection of the
anomaly in the LF, we must have $LFC\leqslant \omega-l_{k}+n$.

From the above analysis, we can identify these ways to
successfully detect intrusions in an LF: (1) making the detector
window larger; (2) making the minimum foreign sequence smaller,
and (3) making the number of MFSs in one locality frame as large
as possible to form a cluster of anomalies
\cite{Hofmeyr98SeqSystemCall}. Since larger detector window will
degrade the efficiency of stide, the latter two options can be
considered to serve as a guideline for choosing proper lengths for
LF and LFC.

\section{Applications}
Apart from strengthening the comprehension of the inherent
dynamics of stide-like AID detectors, the formal framework can
also be applied to accelerate the training of stide-like AID
detectors with less training audit trails, to identify the precise
context of an intrusion and so on. Other than evaluating the
influence of the completeness of training dataset on stide
efficiency, two of its applications will be described in detail:
(1)trimming the normal dataset without losing efficiency for a
given detection performance (thus the training procedure is speed
up), and (2)identifying the intrusion contexts in the intrusive
dataset.

\subsection{Experimental setup and datasets}

\begin{table}[h]
\centering
  \caption{The dataset specifications.}
  \label{tbl:dataSpec}
  \begin{scriptsize}
  \begin{tabular}{|l|l|c|c|}
    \hline
    Normal       & Intrusive & No. of & No. of\\
    Datasets     & Datasets  & Traces & System Calls \\
    \hline
    \hline   live-named-UNM&   ----            & 142 & 9230572 \\
    \hline      ----       & buffer overflow-1 & 3   & 969 \\
    \hline      ----       & buffer overflow-2 & 2   & 831 \\
    \hline
    \hline   live-lpr-MIT  &   ----            & 2703& 2926304 \\
    \hline      ----       & lprcp             & 1001& 165248 \\
    \hline
    \hline   sendmail-CERT &  ----             & 294 & 1576086 \\
    \hline      ----       & syslog-local-1    & 6   & 1516 \\
    \hline      ----       & syslog-local-2    & 6   & 1574 \\
    \hline      ----       & syslog-remote-1   & 7   & 1861 \\
    \hline      ----       & syslog-remote-2   & 4   & 1553 \\
    \hline      ----       & cert-sm565a       & 3   & 275 \\
    \hline     ----        & cert-sm5x         & 8   & 1537 \\
    \hline
    \hline    sendmail-UNM & ----              & 346 & 1799764 \\
    \hline        ----     & decode            & 36  & 3067 \\
    \hline        ----     & forward loops     & 36  & 2569 \\
    \hline        ----     & sunsendmailcp     & 3   & 1119 \\
    \hline
    \hline   syn-wu-ftpd   &    ----           & 8   & 180315 \\
    \hline     ----        & misconfiguration  & 5   & 1363 \\
    \hline

    \hline   syn-xlock-UNM &   ----            & 71  & 339177 \\
    \hline      ----       & buffer overflow-1 & 1   & 489 \\
    \hline      ----       & buffer overflow-2 & 1   & 460 \\
    \hline
  \end{tabular}
  \end{scriptsize}
\end{table}

For the convenience of comparison, the datasets
\cite{Dataset94UNM} that are used in \cite{Warrender99Comparison}
\cite{Kymie03DetermineOpLmts} have been used in our experiments as
well. In addition, we have discarded the normal datasets of
several processes that are too small to use in our framework. The
normal and intrusive datasets for selected processes are specified
in Table~\ref{tbl:dataSpec}. From the table, our selected datasets
represent most processes and intrusions into the processes.
Furthermore, to analyze the characteristics of every intrusion,
its intrusive dataset, even into the same process, is treated as
an individual dataset.

\subsection{The influence of the completeness of training dataset on stide efficiency}
In this section, the influence of the completeness of the training
dataset on the efficiency of stide detectors will be evaluated.
For that purpose, we regard the normal dataset for every process
to be complete for the normal model of the process, and we induce
incompleteness by splitting the normal dataset into a training
dataset  and a test dataset. To remove any dependency, we choose
the training datasets with $m$ varying sizes $Size_{1}, \dots,
Size_{m}$ and n varying starting points $Pos_{1}, \dots, Pos_{n}$
within the length of the normal dataset $\Sigma_{nml}$. To achieve
it, the normal dataset is treated as a continuous ring using wrap
around of the linear dataset. Given any splitting point $Pos_{i}$
and any size $Size_{j}$, the part from $Pos_{i}$ to
$(Pos_{i}+Size_{j})\%|\Sigma_{nml}|$ is selected as
$\Sigma_{trn}(i,j)$, and whatever remains is chosen as the test
dataset $\Sigma_{tst}(i,j)$. Based on $\Sigma_{trn}(i,j)$ and
$\Sigma_{tst}(i,j)$, the completeness of the training dataset will
be evaluated considering stide efficiency.

On the other hand, in stide-like AID techniques, the frequency
information of the events in the training dataset is not utilized,
so trimming the repeated events (or sequences) in the training
dataset is useful to economize the training time without any loss
of efficiency. In the trimming procedure, the \textit{critical
sections} in a normal dataset are identified to produce a compact
training dataset. One of the requirements for the critical section
is that the stide detectors trained by it must be as efficient as
when they are trained by the complete untrimmed dataset. Finally,
the most compact critical section in the normal dataset are chosen
for stide without sacrificing its efficiency.

To achieve it, we also develop two graphical tools to make it easy
and convenient to analyze the completeness of the training
dataset, and the characteristics of the datasets. They are
described below.

\subsubsection{MFS-MSS Average Curves}
These curves are inspired by Theorem~\ref{theorem:Tendency}, which
can be used to depict the influence of the completeness of the
training dataset on the detection efficiency graphically. At the
same time, our objective for these curves is to evaluate the
dynamics in stide efficiency with the completeness of training
dataset, thus, we only concern about the size of training dataset.
For this reason, the average values for
$|MSS|_{min}(\Sigma_{tst}|\Sigma_{trn})$ and
$|MFS|_{min}(\Sigma_{int}|\Sigma_{trn})$ for a given training data
size $Size_{j}\;(1\leqslant j\leqslant m)$ are first calculated:
\begin{eqnarray}
&&\hspace{-20pt}|MSS_{min}|_{avg}(j)\nonumber\\
&=&\frac{1}{n}*\mathop{\sum_{i=1}^{n}}|MSS|_{min}(\Sigma_{tst}(i,j)| \Sigma_{trn}(i,j)) \\
&&\hspace{-20pt}|MFS_{min}|_{avg}(j)\nonumber\\
&=&\frac{1}{n}*\mathop{\sum_{i=1}^{n}}|MFS|_{min}(\Sigma_{int}
|\Sigma_{trn}(i,j))
\end{eqnarray}
Then, we plot the average values of
$|MSS|_{min}(\Sigma_{tst}|\Sigma_{trn})$ and
$|MFS|_{min}(\Sigma_{int}|\Sigma_{trn})$ against the corresponding
sizes of the dataset $\Sigma_{trn}$. We call the resulting graphs
as \textbf{MFS-MSS Average Curves (MMAC)}
(Figure~\ref{fig:MSSMFSCurve}).

\subsubsection{MFS-MSS Matrix}
Let us first introduce a new concept `critical section'. Within
context of stide, for a splitting point $Pos_{i}$, $Size_{j}$ is a
\textbf{critical section} $CS(i,\lambda)$ if
$|MSS|_{min}(\Sigma_{tst}(i,j)|\Sigma_{trn}(i,j))\geq\lambda$ but
$|MSS|_{min}(\Sigma_{tst}(i,j-1)|\Sigma_{trn}(i,j-1))<\lambda$.
Obviously, the critical section is indispensable to provide stide
detectors with the detection performance $\lambda$. Other than the
critical section $CS(i,\lambda)$, the remaining part of the normal
dataset can be discarded as it has negligible effect on the stide
detection efficiency.

The MFS-MSS Matrix (MMM) is defined in order to help identify the
critical sections in the normal dataset with respect to the
predefined detection performance $\lambda$. In the matrix, the
columns (the horizontal axis) are defined by the splitting sizes
of the training dataset $\{Size_{1}, Size_{2}, \dots, Size_{m}\}$,
and the rows (the vertical axis) are defined by the splitting
points of the training dataset $\{Pos_{1}, Pos_{2}, \dots,
Pos_{n}\}$ (as in Figure~\ref{fig:MMM}). According to our proposed
formal framework (especially from Eqn~(\ref{eqn:IncompleteMFS})
and Theorem~\ref{theorem:Tendency}), an entry $MMM(i,j)$ in an MMM
matrix will be labeled as `\textbf{efficient}' if
$|MSS|_{min}(\Sigma_{tst}(i,j)| \Sigma_{trn}(i,j))\geqslant
\lambda$, otherwise, it is labeled as `\textbf{inefficient}'.
Furthermore, for every specific pair of $Pos_{i}$ and $Size_{j}$,
if $MMM(i, j)$ is \textit{inefficient} but $MMM(i,j+1)$ is
\textit{efficient}, the \textit{transition} from the inefficient
entry $MMM\\(i, j)$ to the efficient entry $MMM(i,j+1)$ is named
as an \textbf{efficiency transition} in the MMM matrix. From the
efficiency transition, it can be concluded that the section in
$\Sigma_{nml}$ from $Pos_{i}$ to
$(Pos_{i}+Size_{j+1})\%|\Sigma_{nml}|$\footnote{\ It is done in a
wrap-round fashion as the normal dataset splitting policy in the
same application.} is critical for building efficient stide
detectors, i.e., it is a \textit{critical section}
$CS(i,\lambda)$.

After identifying the critical sections for all splitting points,
we choose the \textbf{most compact critical section}
$MCCS(\lambda)(=CS(i,\lambda))$ in the normal dataset as the
training dataset for stide. As its name implies, for any other
critical section $CS(k,\lambda)$ ($i\neq k$), $|MCCS(\lambda)|\leq
|CS(k,\lambda)|$. Since the redundant parts in the normal dataset
can be trimmed by using $MCCS(\lambda)$, the training time for the
stide detectors can be substantially reduced without sacrificing
the detection performance. As an added benefit, the size of
$MCCS(\lambda)$ in the normal dataset provides an intuitive
measure of the complexity of a process. This is because,
intuitively, with respect to the same detection performance, the
more complex the process is, the larger $MCCS(\lambda)$ is.
Furthermore, this technique for dataset trimming can be utilized
in other domains as well, such as information retrieval and
computer forensic.

\paragraph{Effect of the trimming scheme}
As mentioned earlier, in order to be valid, any trimming of the
training dataset must not lead to any loss in the efficiency of
stide detectors. That our trimming procedure indeed satisfies the
criterion is shown by the Theorem~\ref{theorem:whyTrim}.

\begin{theorem}
\label{theorem:whyTrim} Let $\Sigma_{trn}^{cs}$ denote the
critical section, and $\Sigma_{tst}^{cs}$ the remaining part in
the normal dataset. Thus, $\Sigma_{trn}^{cs}\odot
\Sigma_{tst}^{cs}=\Sigma_{nml}$. The future normal dataset is
denoted as $\Sigma_{new}$. Then, for all (known and unknown)
intrusions with
$MFS(\Sigma_{int}|\Sigma_{nml}\odot\Sigma_{new})\leqslant
\lambda$,
\begin{eqnarray*}
&&\hspace{-20pt}|MSS|_{min}(\Sigma_{new}|\Sigma_{nml})\geqslant
|MFS|_{min}(\Sigma_{int}|\Sigma_{nml}\odot\Sigma_{new})\\
&&\Rightarrow
|MSS|_{min}(\Sigma_{tst}^{cs}\odot\Sigma_{new}|\Sigma_{trn}^{cs})\\
&&\hspace{20pt}\geqslant|MFS|_{min}(\Sigma_{int}|\Sigma_{nml}\odot\Sigma_{new})
\end{eqnarray*}
\end{theorem}

\begin{proof} In the trimming scheme, we assume that
$|MSS|_{min}$\\ $(\Sigma_{tst}^{cs}|\Sigma_{trn}^{cs})=\lambda$
($>0$), i.e., the detection performance of the stide detector
built by the critical section is stable at $\lambda$ to detect
intrusions that cause MSSs smaller than $\lambda$. From its
definition,
$|MSS|_{min}(\Sigma_{tst}^{cs}\odot\Sigma_{new}|\Sigma_{trn}^{cs})$
is affected by the foreign sequence(s) $S$ ($S\in
MFS_{min}(\Sigma_{tst}^{cs}\odot\Sigma_{new}|\Sigma_{trn}^{cs})$)
under the following two scenarios:\\
\noindent\textbf{CASE 1:}
$S\in SS(\Sigma_{tst}^{cs})$;
\begin{eqnarray*}
&&\hspace{-20pt}S\in SS(\Sigma_{tst}^{cs}), S\not\in
SS(\Sigma_{trn}^{cs})\\
&&\hspace{-20pt}\Rightarrow |S|=\lambda+1\\
&&\hspace{-20pt}\Rightarrow|MSS|_{min}(\Sigma_{tst}^{cs}\odot\Sigma_{new}|\Sigma_{trn}^{cs})= \lambda\\
&&\hspace{-20pt}\Rightarrow|MSS|_{min}(\Sigma_{tst}^{cs}\odot\Sigma_{new}|\Sigma_{trn}^{cs})\\
&&\hspace{20pt}\geqslant|MFS|_{min}(\Sigma_{int}|\Sigma_{nml}\odot\Sigma_{new})
\end{eqnarray*}
\noindent\textbf{CASE 2:} $S\not\in SS(\Sigma_{tst}^{cs})$;
\begin{eqnarray*}
&&\hspace{-20pt}S\not\in SS(\Sigma_{tst}^{cs}),S\not\in SS(\Sigma_{trn}^{cs})\\
&&\hspace{-20pt}\Rightarrow S\subset\Sigma_{new},S\not\subset\Sigma_{nml}\\
&&\hspace{-20pt}\Rightarrow |MSS|_{min}(\Sigma_{new}|\Sigma_{nml})= |S|-1\\
&&\hspace{-20pt}\Rightarrow
|MSS|_{min}(\Sigma_{new}|\Sigma_{nml})= |MSS|_{min}(\Sigma_{tst}^{cs}\odot\Sigma_{new}|\Sigma_{trn}^{cs})\\
&&\hspace{-20pt}\Rightarrow
|MSS|_{min}(\Sigma_{tst}^{cs}\odot\Sigma_{new}|\Sigma_{trn}^{cs})\\
&&\hspace{20pt}\geqslant|MFS|_{min}(\Sigma_{int}|\Sigma_{nml}\odot\Sigma_{new})
\end{eqnarray*}
Based on the results under these two scenarios, the theorem is
proved.
\end{proof}

\subsection{Experimental evaluations}
\begin{figure*}[t]
\centering
\begin{subfigure} [Process `named'.]
{\epsfig{file=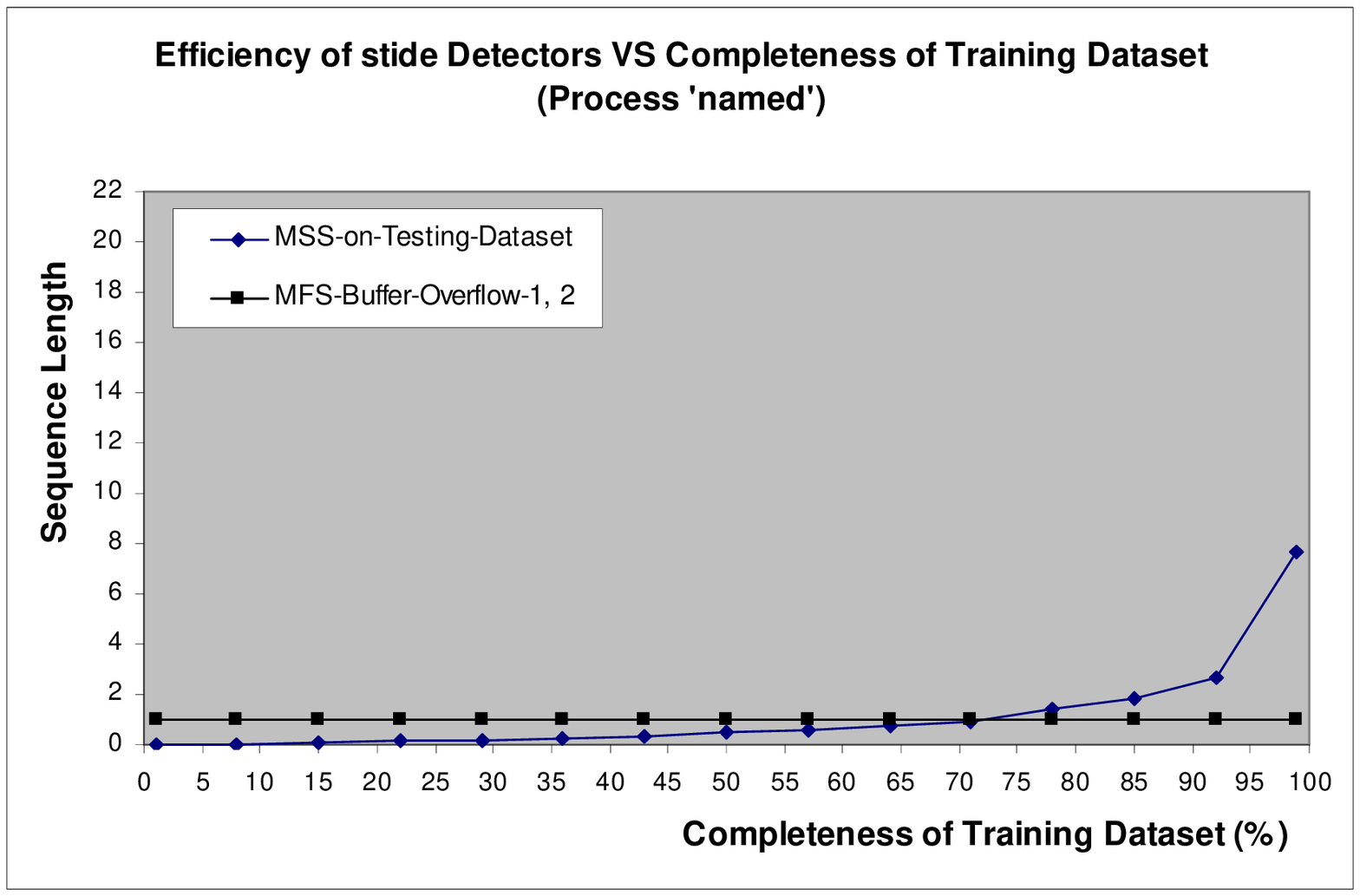, height=1.2in, width=2.2in} }
\end{subfigure}
\centering
\begin{subfigure} [Process `lpr' from MIT.] {\epsfig{file=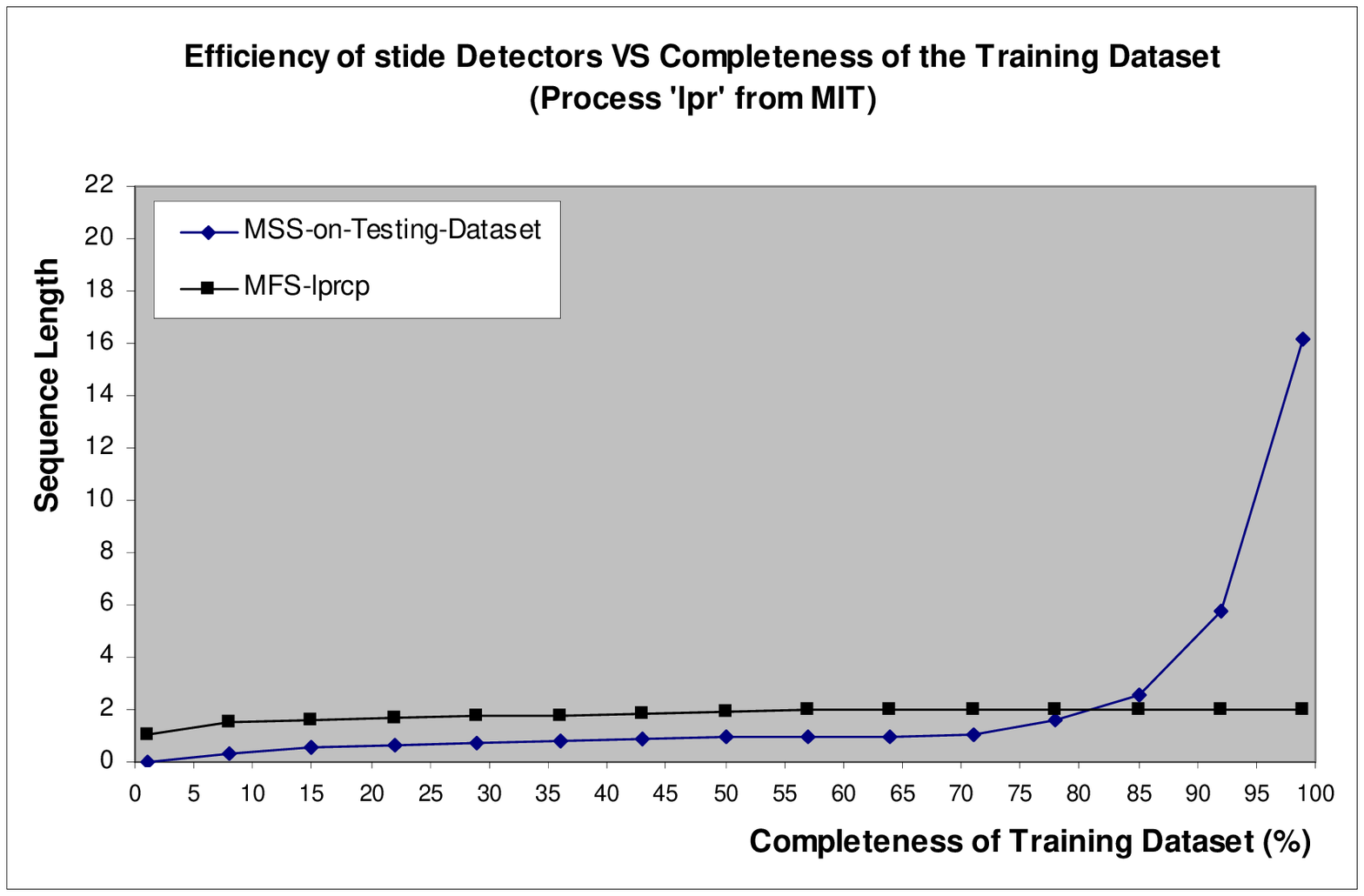, height=1.2in, width=2.2in}}
\end{subfigure}
\begin{subfigure} [Process `sendmail' from CERT.]
{\epsfig{file=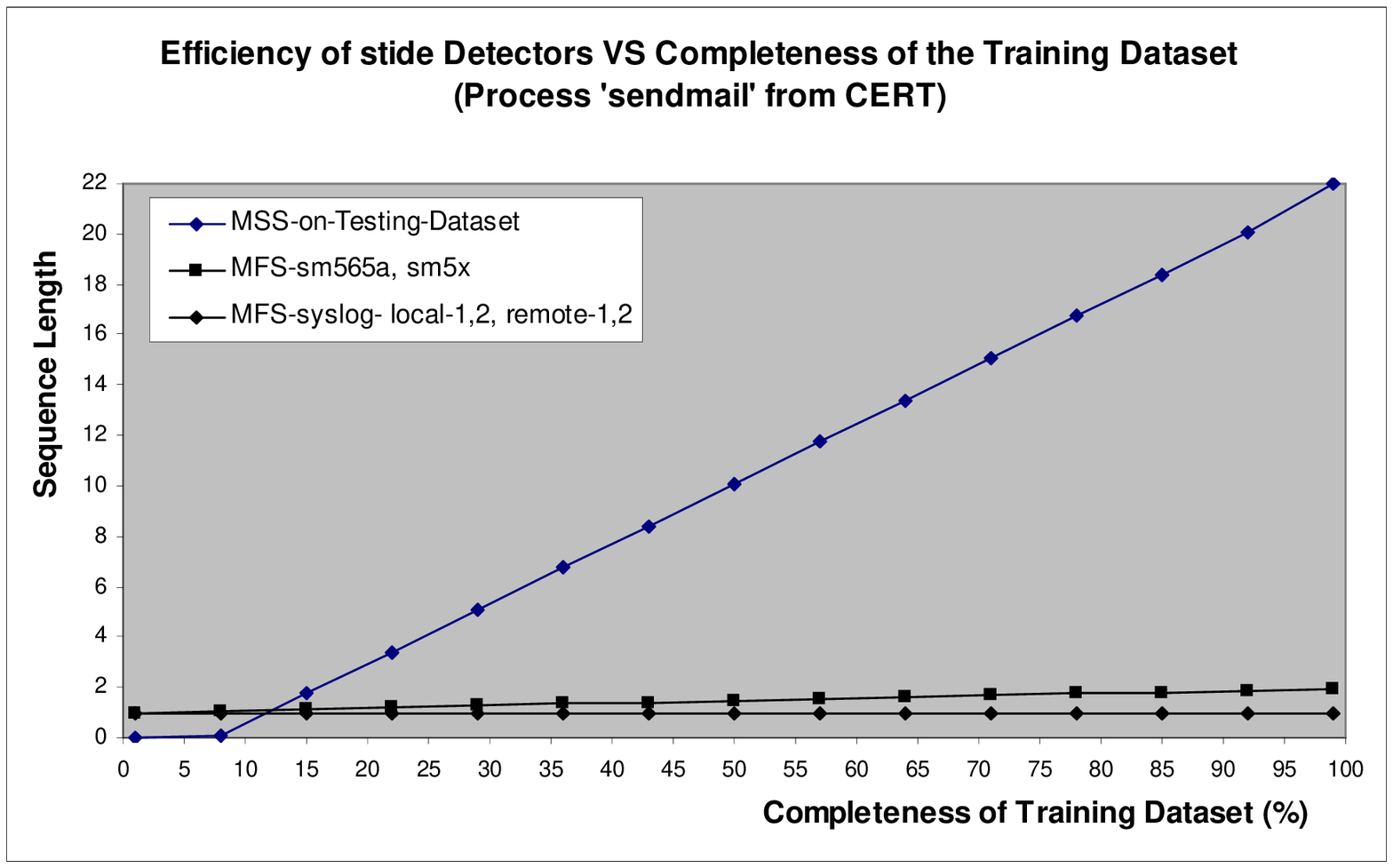, height=1.2in, width=2.2in}}
\end{subfigure}
\begin{subfigure} [Process `sendmail' from UNM.] {\epsfig{file=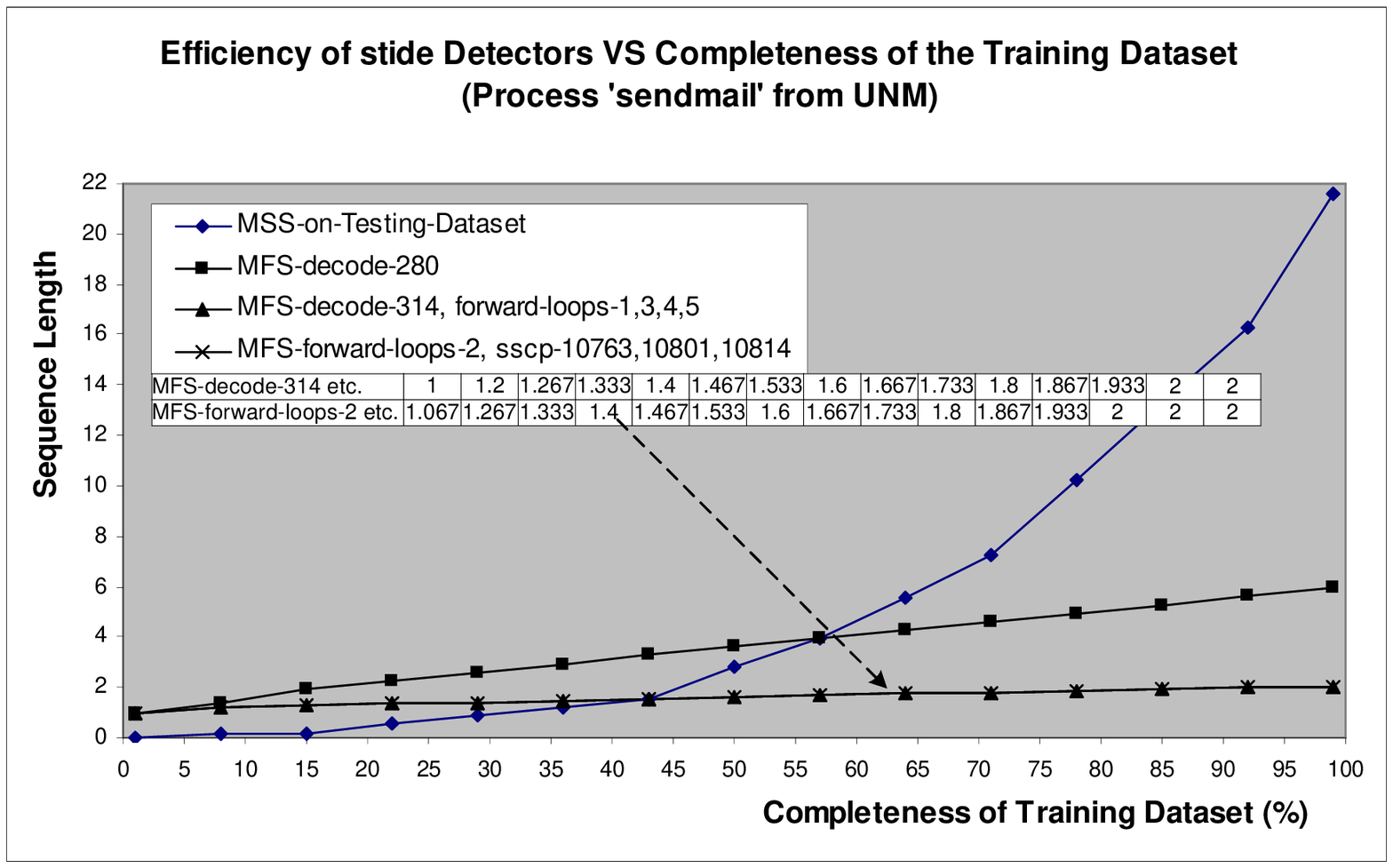, height=1.2in, width=2.2in}}
\end{subfigure}
\begin{subfigure} [Process `ftpd'.]
{\epsfig{file=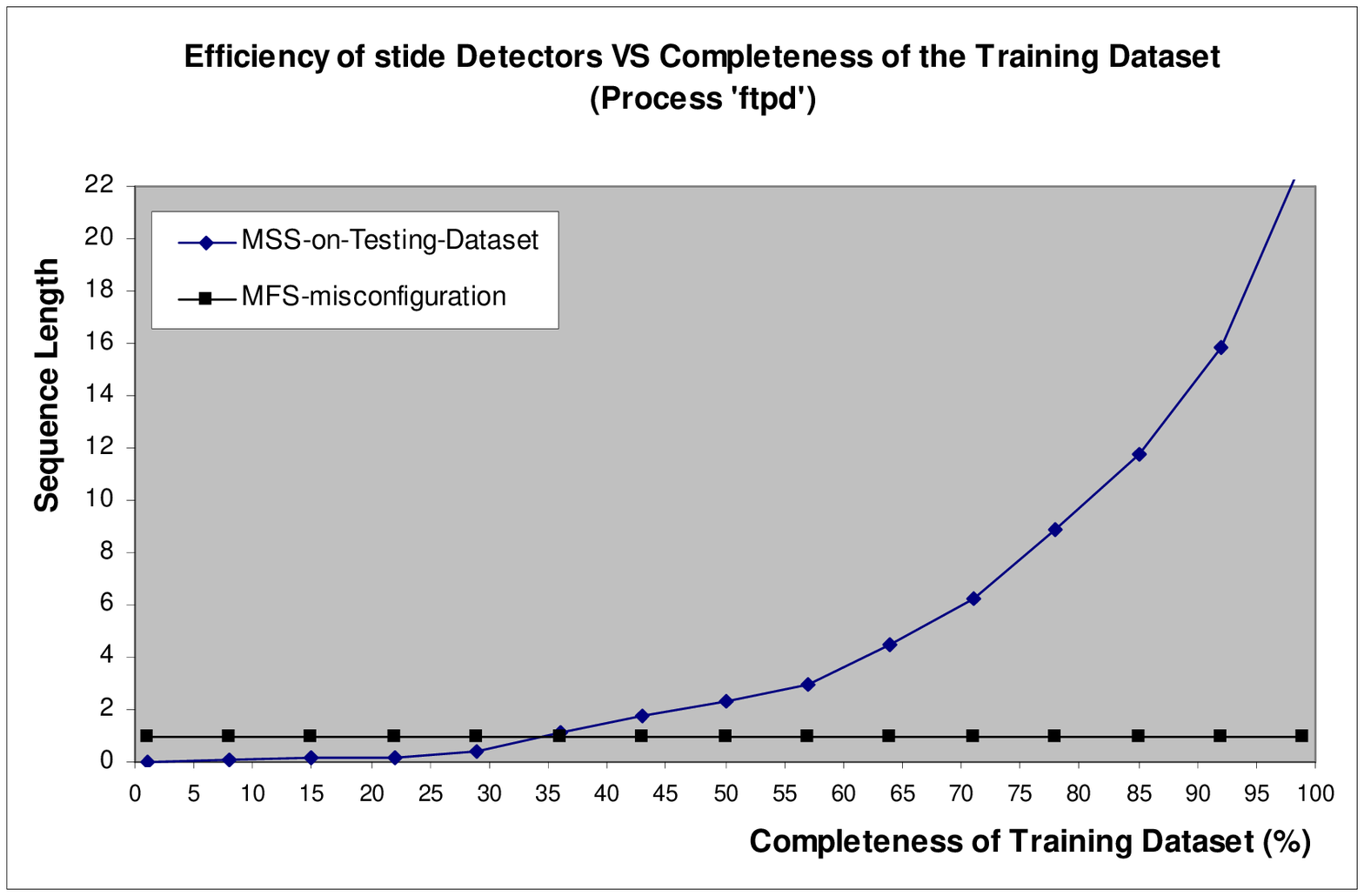, height=1.2in, width=2.2in}}
\end{subfigure}
\begin{subfigure} [Process `xlock'.]
{\epsfig{file=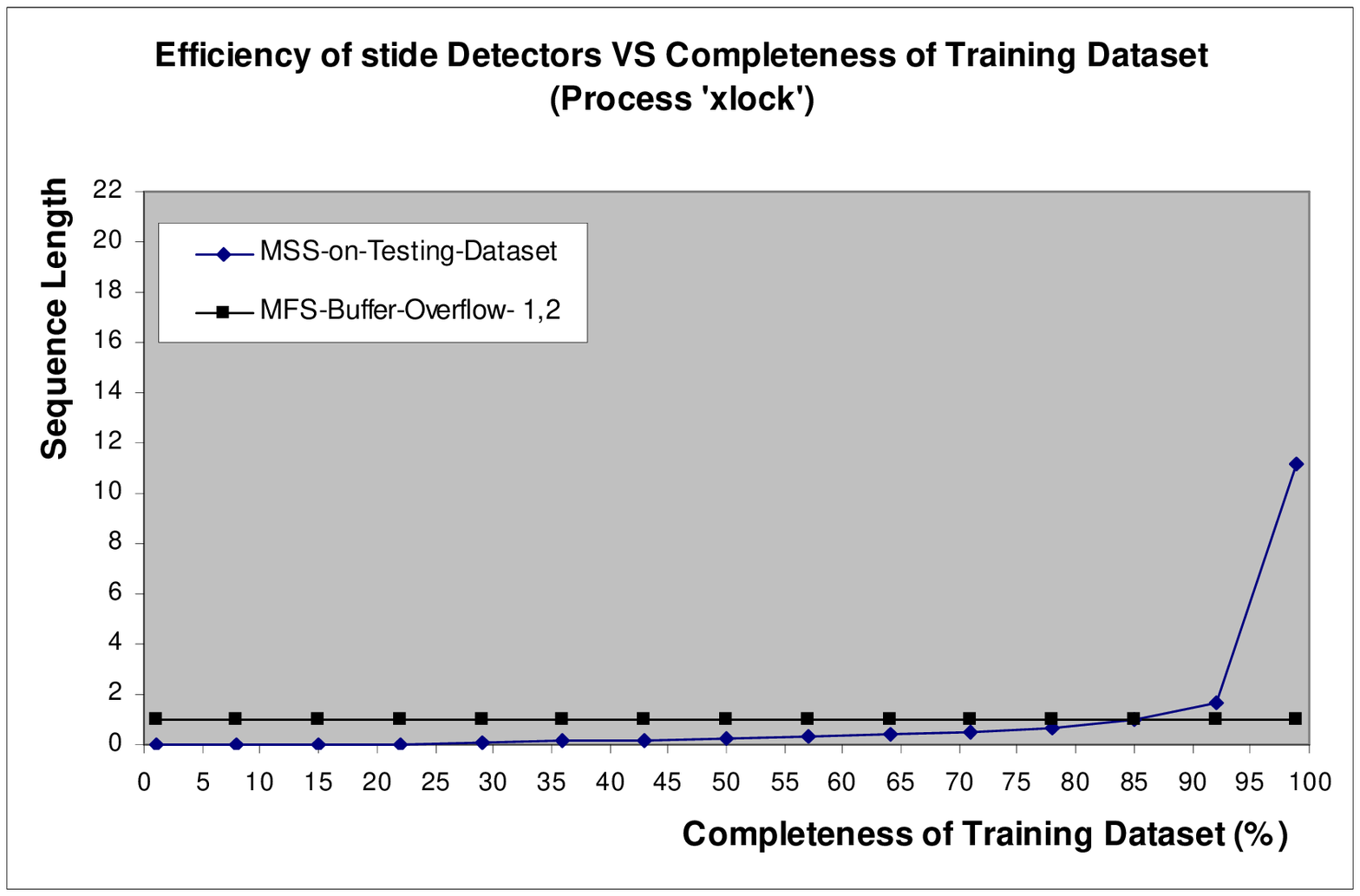, height=1.2in, width=2.2in}}
\end{subfigure}
\caption{MFS-MSS average curves for different processes.}
\label{fig:MSSMFSCurve}
\end{figure*}

The splitting procedure in our application works as follows. The
length of the training dataset $Size_{j}$ is varied from 1 to 99\%
of the normal dataset, with a step of 7\%, and the remaining
portion of the normal dataset is designated as the test dataset.
The splitting position $Pos_{i}$ is also varied dynamically from
1\% to 99\% of the normal dataset, using wrap around, with a step
of 7\%. Thus, $m=n=15$. The maximum length for MSSs in any test
dataset is kept fixed at N=25 (as all the MSSs and MFSs obtained
are well within this limit).

In our experiments, the following aspects of the framework will be
evaluated:
\begin{enumerate}
\item[A)] The influence of the completeness of the training
dataset on the MFSs in the intrusive dataset;

\item[B)] The influence of the completeness of the training
dataset on the MSSs in the test dataset;

\item[C)] The effectiveness of the trimming procedure, and the
related graphical tools.
\end{enumerate}

\subsubsection{Evaluating the completeness of the training
dataset} \label{sec::MMAC}
In Figure~\ref{fig:MSSMFSCurve}, the MFS-MSS average curves for
processes are illustrated\footnote{For clarity, in this figure, we
have grouped the intrusions which have the same MFS sequences with
the increase of the completeness of the training dataset. For the
same reason, some MFS sequences will be organized in a table if
they are too near to be distinguishable from their curves, such as
the table in Figure~\ref{fig:MSSMFSCurve}.d.}. From these curves,
the varying sensitivity of the stide detectors to the completeness
of the training dataset is obvious. For processes `named', 'xlock'
and `lpr' from MIT, the efficiency of stide detectors is quite
sensitive to the completeness of the training dataset. For process
`sendmails' from CERT, the efficient stide detectors can be
obtained even with a small size of the training dataset. At the
same time, the minimum foreign sequences of the intrusive dataset
are not affected that much by the completeness of the training
dataset since
$|MFS|_{min}(\Sigma_{int}|\Sigma_{trn}\odot\Sigma_{tst})$ is small
(\emph{less than 2}) for `named', `lpr' and `sendmail' from CERT
(Equation \ref{eqn:IncompleteMFS}). However, if the MFS(s) of an
intrusion is larger than 2, such as 'decode-280', the influence of
the minimum common false positive sequences in the intrusive
dataset can be observed clearly as the completeness of the normal
dataset increases. It is worth noting that the answer to the `Why
6?' \cite{Tan01Why6} question is provided explicitly by
Figure~\ref{fig:MSSMFSCurve}.d.

General speaking, the degree of sensitivity of the stide detectors
to the completeness of the training dataset may be influenced by
the complexity of the processes and/or the audit trails collection
tools. If the function of a process is simple, the complete
training dataset is easy to collect, and the detection efficiency
is more influenced by the intrusion characteristics
$|MFS|_{min}(\Sigma_{int}|\Sigma_{trn}\odot\Sigma_{tst})$
(Theorem~\ref{theorem:minMFPS-MFS}). Otherwise, the completeness
of the training dataset is hard to be guaranteed, and the stide
detector is more sensitive to the completeness of the training
dataset. In other words, stide is very efficient to detect
intrusions into a process with simple function, but its efficiency
will be deteriorated when detecting a complex process. For
instance, stide is not appropriate to detect the intrusions in
Internet since the traffic behaviors in Internet are very dynamic,
and even evolved with time.

\subsubsection{Identifying critical sections using MMM}
\label{sec:MMM}
\begin{figure*}[t]
\centering
\begin{subfigure} [Process `named'.]
{\epsfig{file=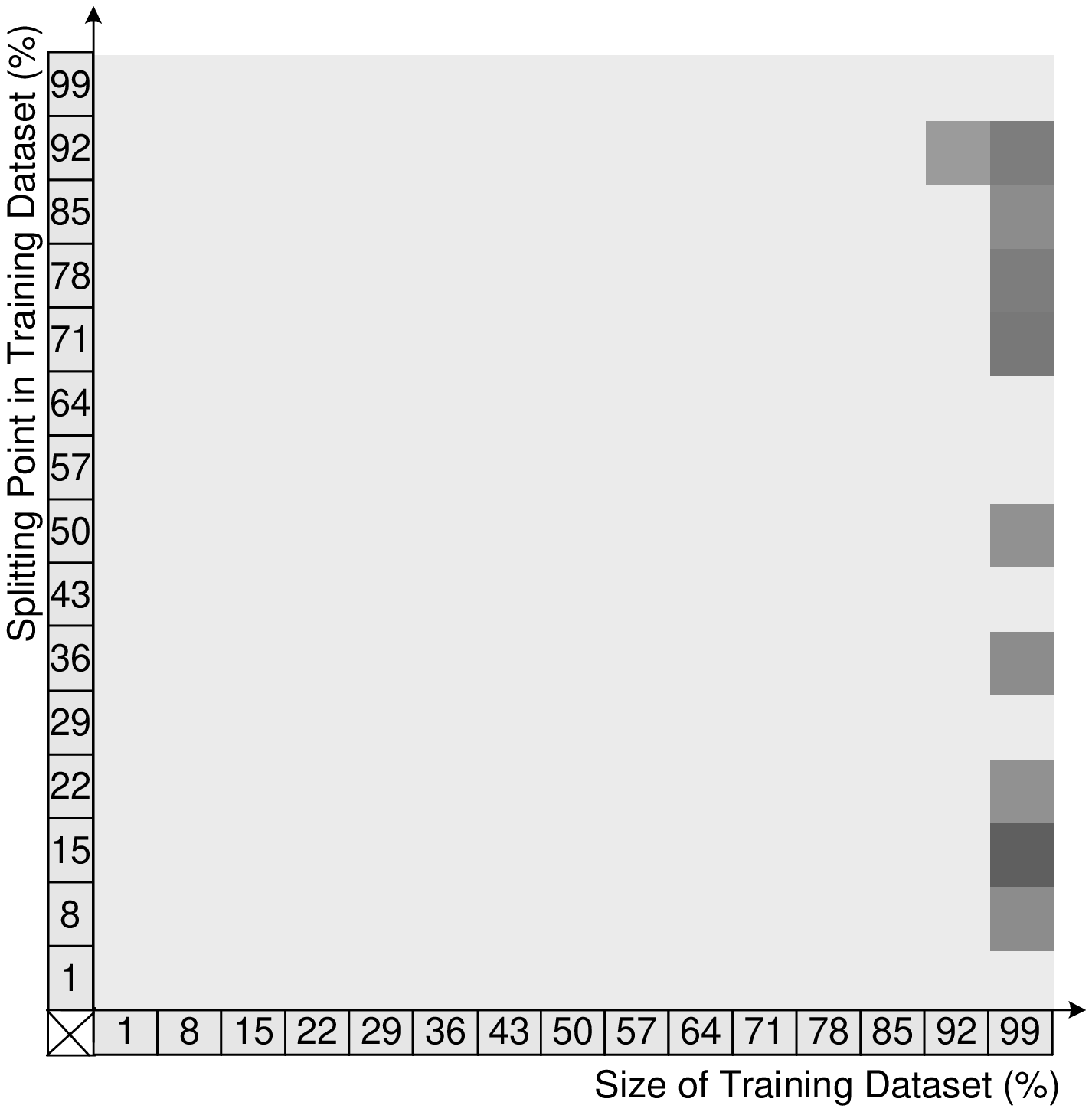, width=1.0in, height=1.0in} }
\end{subfigure}
\begin{subfigure} [Process `lpr'.]
{\epsfig{file=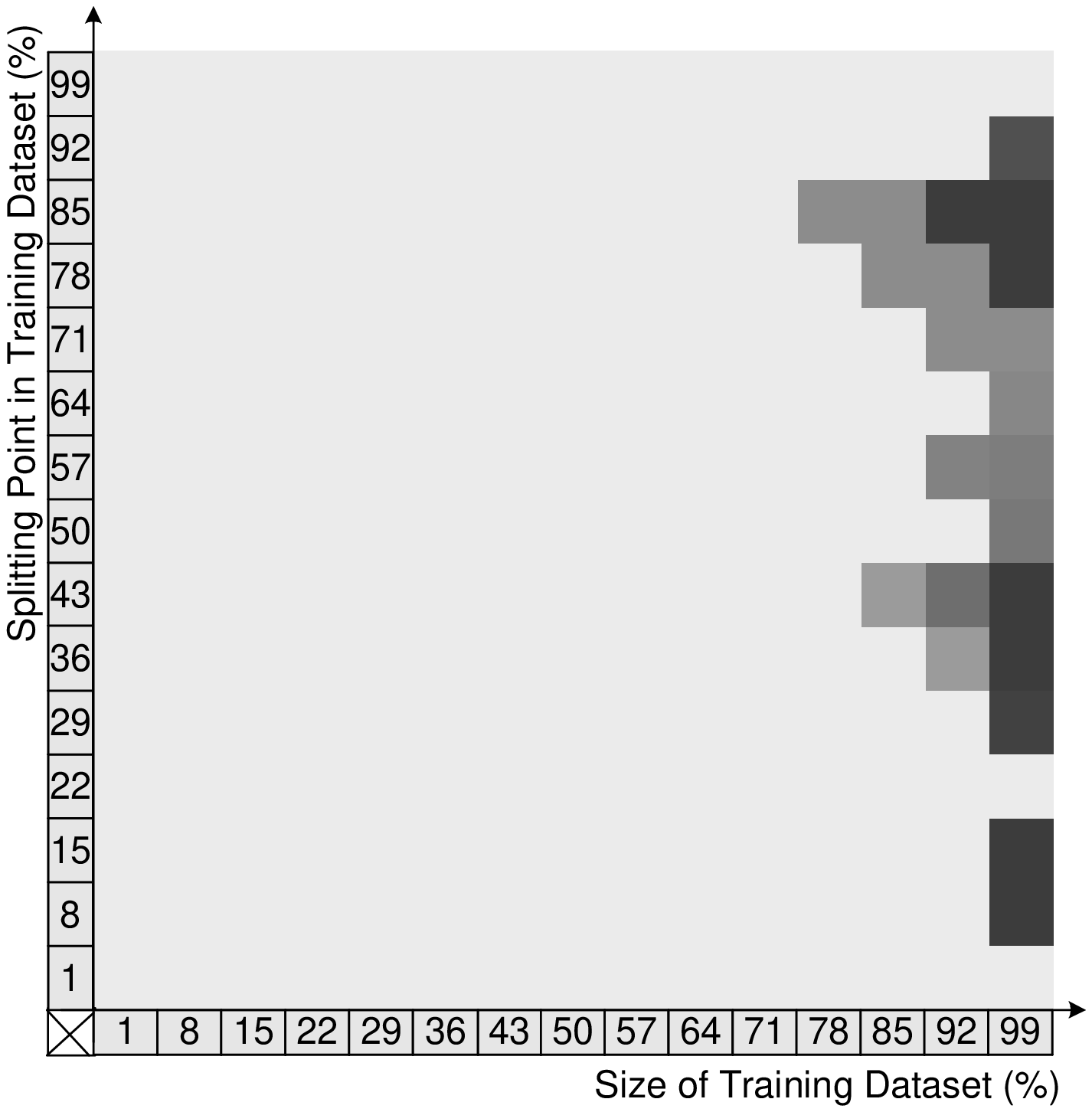, width=1.0in, height=1.0in}}
\end{subfigure}
\begin{subfigure} [Process `sendmail' from CERT.]
{\epsfig{file=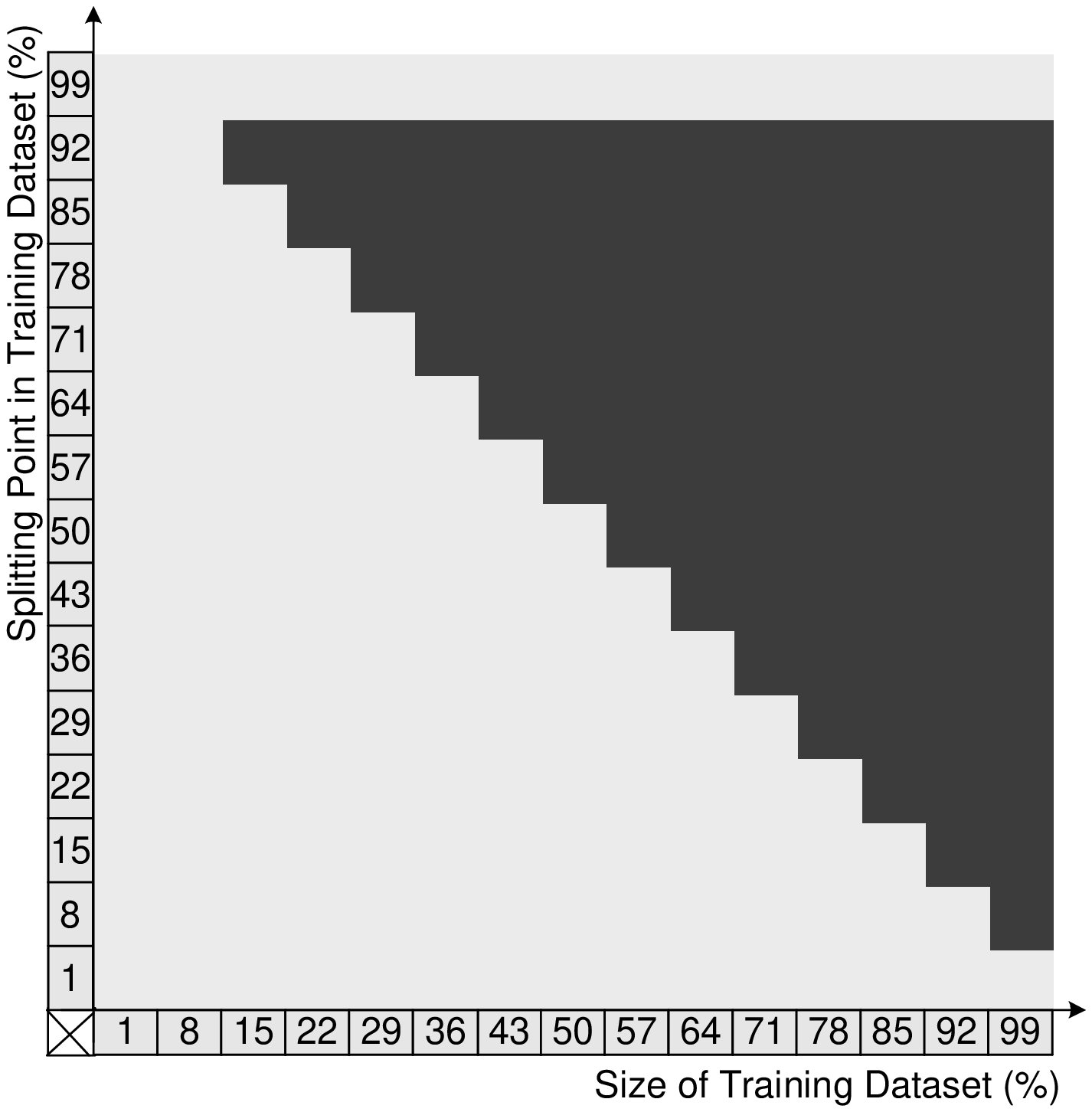, width=1.0in, height=1.0in}}
\end{subfigure}
\begin{subfigure} [Process `sendmail' from UNM.]
{\epsfig{file=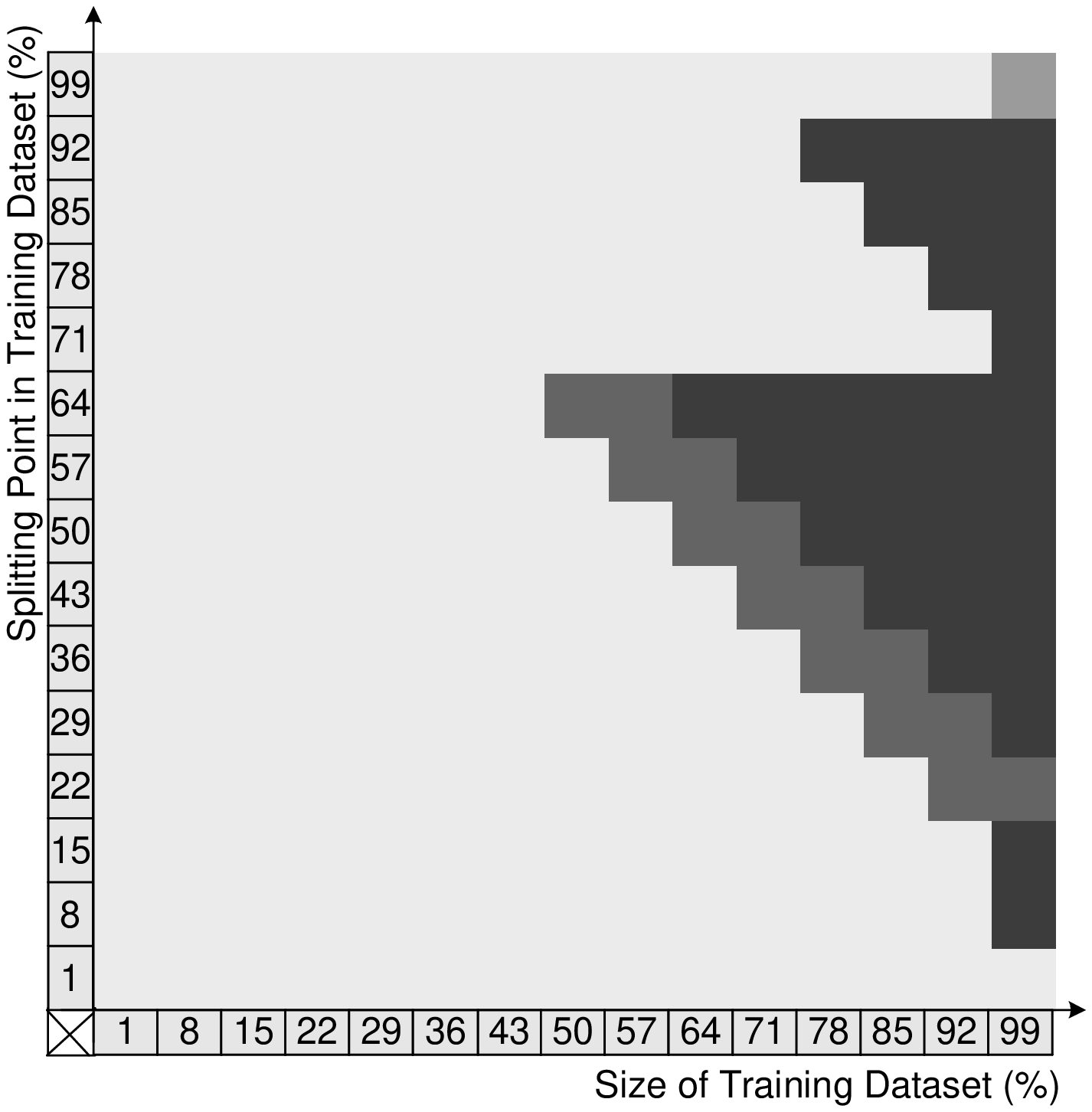, width=1.0in, height=1.0in}}
\end{subfigure}
\begin{subfigure} [Process `ftpd'.]
{\epsfig{file=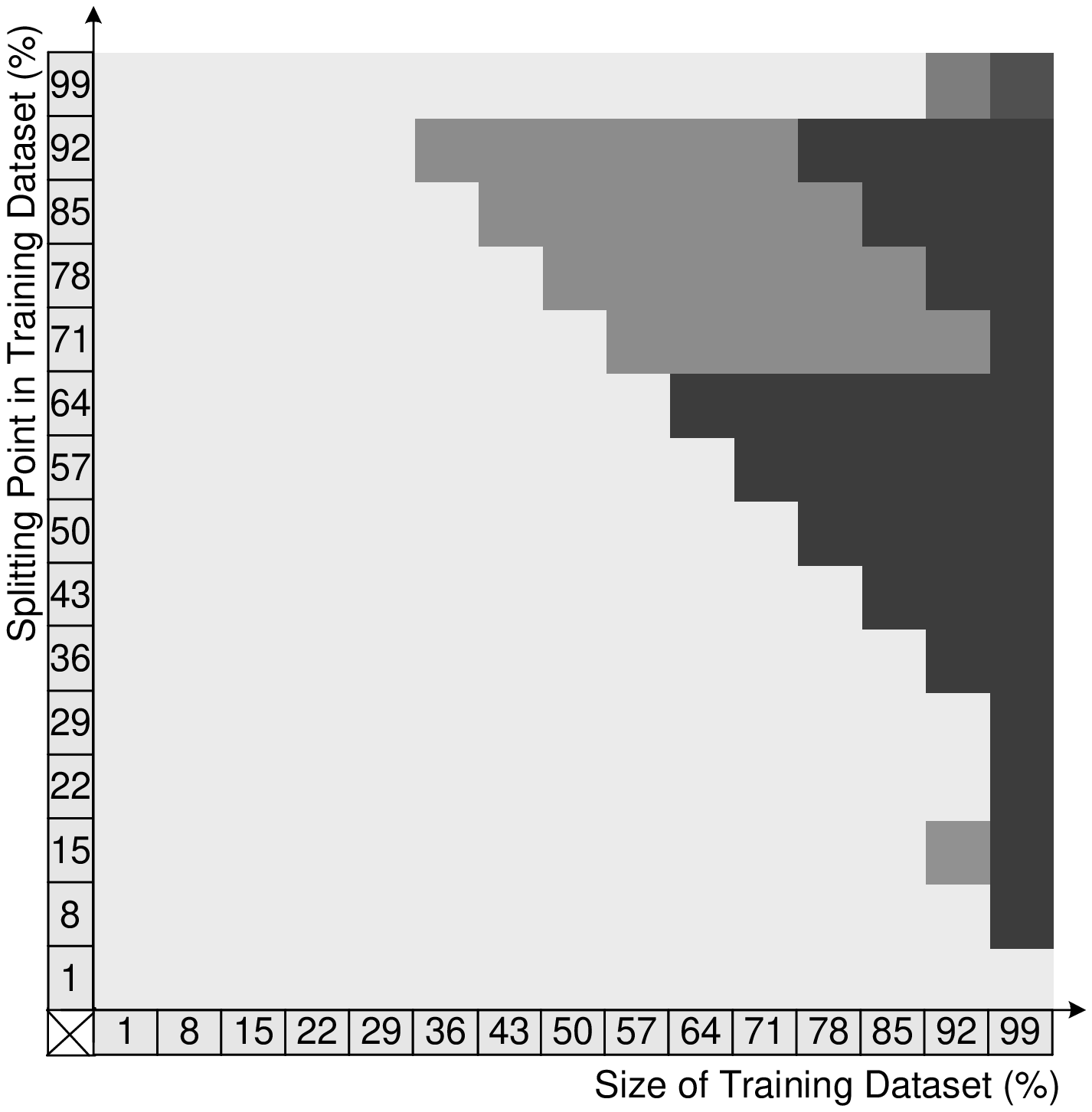, width=1.0in, height=1.0in} }
\end{subfigure}
\begin{subfigure} [Process `xlock'.]
{\epsfig{file=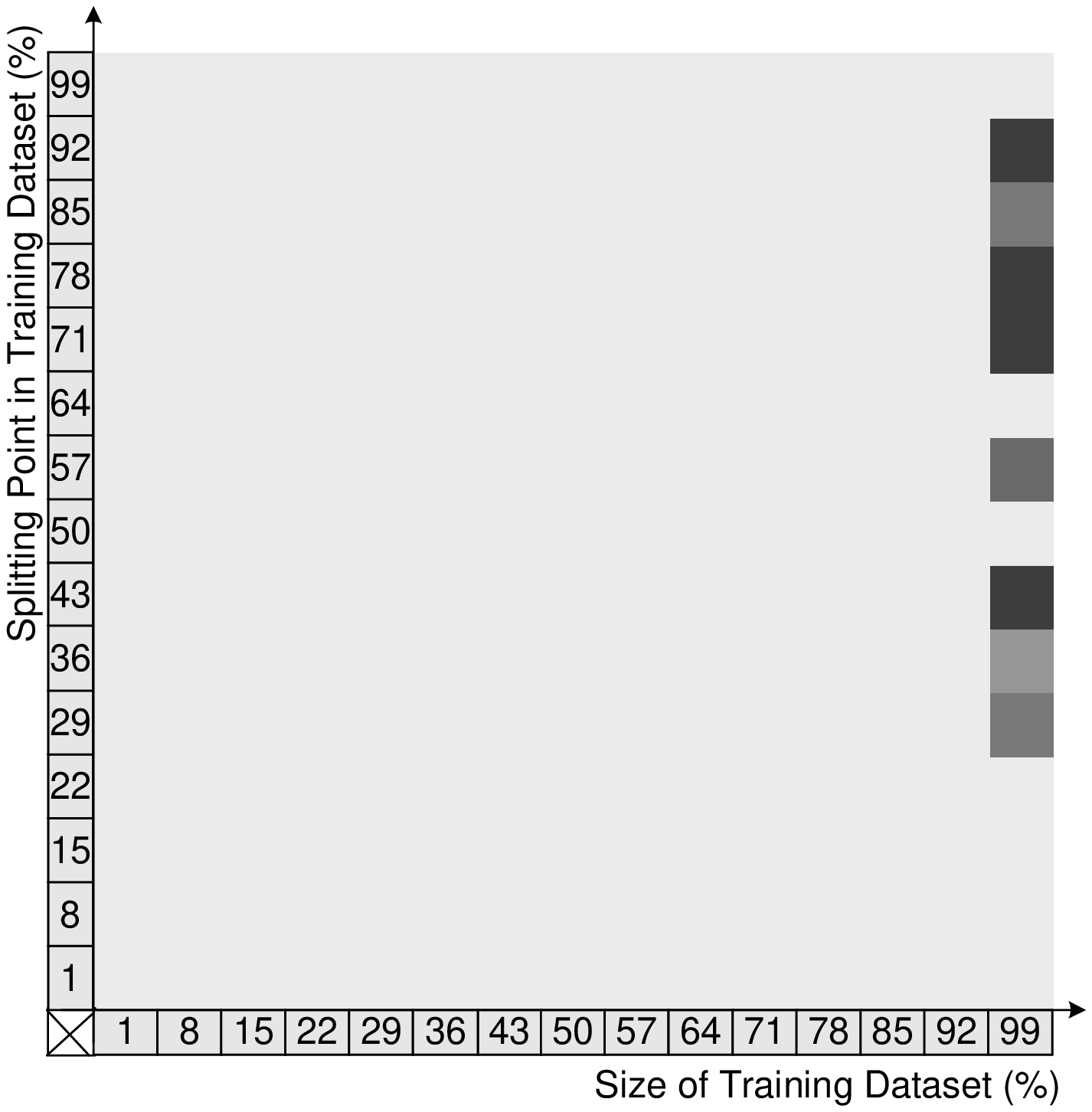, width=1.0in, height=1.0in} }
\end{subfigure}
\caption[center]{MFS-MSS Matrix for different processes
($\lambda=6$).} \label{fig:MMM}
\end{figure*}
In the MMM matrix, the efficiency of every entry is indicated by
its darkness, which is defined by the value $|MSS|_{min}$\\
$(\Sigma_{tst}(i,j)|\Sigma_{trn}(i,j))-\lambda$. As in
Figure~\ref{fig:MMM}, the darker elements of the matrix indicate
the \emph{efficient entries}, and the lighter elements of the
matrix indicate the \textit{inefficient entries}. Therefore, based
on Theorem~\ref{theorem:Tendency}, the darker the entry in the MMM
matrix is, the more possibility to train efficient stide detectors
from the training dataset determined by the entry (i,j).
Furthermore, for every splitting point, the efficiency transition
is clearly visible as it is the transition from a lighter entry to
the \textit{first} darker one.

In our experiments, we let $\lambda=6$. From these efficiency
transitions in the MMM matrices (Figure~\ref{fig:MMM}), the
\begin{table}[h]
\centering \caption{The most compact critical sections when
$\lambda=6$.}\label{tbl:MCCS}
\begin{scriptsize}
\begin{tabular}{|p{1.135in}|c|l|}
  \hline Process             & $MCCS(\lambda)$(\%)            &$|MCCS(\lambda)|$\\
  \hline `named' from UNM    & [92, 100]$\cup$[0, 84] &92\% - - 8492126\\
  \hline `lpr' from MIT      & [85, 100]$\cup$[0, 63] &78\% - - 2282517\\
  \hline `sendmail' from CERT& [92, 100]$\cup$[0, 7]  &15\% - - 236412\\
  \hline `sendmail' from UNM & [64, 100]$\cup$[0, 14] &50\% - - 899882\\
  \hline `wu-ftpd' from UNM  & [92, 100]$\cup$[0, 28] &36\% - - 64913\\
  \hline `xlock' from UNM    & [71, 100]$\cup$[0, 70] &99\% - - 335785\\
  \hline
\end{tabular}
\end{scriptsize}
\end{table}
critical sections in the normal dataset for any process can be
identified easily for every splitting point. For example, for the
process `lpr', at the splitting point $Pos_{12}=78\%$,
$CS(12,6)=[78\%,100\%]\cup[0,63\%]$. Finally, the most compact
critical section is gotten for every process (e.g., for `lpr',
$MCCS(6)=[85\%,100\%]\cup[0,63\%]$). In our experiments, the MCCSs
for various processes in the normal datasets are shown in Table
\ref{tbl:MCCS}. Obviously, the beginning part $[0,7\%]$ and the
end part $[92\%,100\%]$ are included in all these most compact
critical sections. As discussed in \cite{Li04ACI}, the beginning
and end transactions of a process are critical in building the
normal behavior model, and thus affect the stide efficiency.

From the critical section set for every normal dataset in MMM
matrix, the sensitivity of the efficiency of stide detectors to
the constitution and completeness of the training dataset can be
identified more meticulously. Furthermore, from the actual size
(i.e., not the percentage) of $MCCS(\lambda)$ in the normal
dataset of a process, we can get a rough indication of the
complexity of the process. For example, from Table \ref{tbl:MCCS},
we can infer the following order in terms of complexity among
various processes in the experimental datasets: ``\textit{wu-ftpd
$\leqslant$ sendmail from CERT $\leqslant$ xlock $\leqslant$
sendmail from UNM $\leqslant$ lpr from MIT $\leqslant$ named}".

\subsection{Identifying the intrusion context}
\label{Application2}
By identifying the context associated with each alarm generated by
the stide detector, it is possible to separate true alarms from
false alarms. This then can be useful in designing more accurate
detectors, and reducing (or removing) the false alarms.


\subsubsection{Foreign sequences graphs}
To identify the intrusion context, the intrusion dataset can be
processed in two ways: (1) splitting it into blocks and evaluating
every block; (2) evaluating every event in the intrusive dataset.
We use the second option because the splitting process has the
potential to break the foreign sequences, producing spurious and
misleading anomalous sequences. To evaluate the impact of every
event $e_{i}$, we determine the \textit{Foreign Sequence Length}
of $e_{i}$ or $FSL(e_{i})$, which is the length of the first
precedent foreign sequence that ends with $e_{i}$
(Algorithm.\ref{alg:FSL}). We plot the values of $FSL(e_{i})$
against the index $i$ of the event to generate a graph called the
\textbf{foreign sequence graph (FSG)}.
\begin{algorithm}[h]
\caption{Calculating the foreign sequence lengths for all the
events in an intrusive dataset.}\label{alg:FSL}
\begin{scriptsize}
\begin{algorithmic}
\REQUIRE The event sequence of the process $e_{1}, e_{2}, \dots,
e_{p}$; and the stide self model series $stide_{1}, stide_{2},
\dots, stide_{N}$;
\end{algorithmic}
\begin{algorithmic}
\FOR{i=1 to p} \STATE seq=NULL; $FSL(e_{i})=N+1$; \FOR{j=0 to N-1}
\STATE If i-j$\leqslant$0 break; \STATE Insert $e_{i-j}$ to seq as
the first element; \IF{seq is-not-in stide(j+1)} \STATE
$FSL(e_{i})=length(seq)$; break;\ENDIF \ENDFOR \ENDFOR
\end{algorithmic}
Output $FSL(e_{1}), FSL(e_{2}), \dots, FSL(e_{p})$
\end{scriptsize}
\end{algorithm}

According to our proposed framework, the MFSs can be determined
from the foreign sequence graph as follows. In practice, if an MFS
in one intrusive process is $e_{a}e_{a+1}\dots e_{b}$
($b-a+1\leqslant N$), it will be identified by the lowest point in
its foreign sequence graph, which is expressed as
$FSL(e_{b})=b-a+1$, since $FSL(e_{b-1})\geqslant b-a+1$, and
$FSL(e_{b+1})\geqslant b-a+1$. Thus, the precedent $b-a+1$ events
constitute the intrusion context, that is, the minimum foreign
sequence.

\subsubsection{Minimum foreign sequences in FSGs}
According to the definition of minimum foreign sequences, the
non-MFS foreign sequences are formed by augmenting minimum foreign
sequences. For example, suppose that an MFS is
$x_{1},x_{2},\dots,x_{l}$, the foreign sequence can be constituted
as follows: $y_{1}\dots y_{m} x_{1}x_{2}\dots x_{l} y_{m+1}\dots
y_{m+n}$, where, $m\geqslant 0$, $n\geqslant 0$, $y_{1}\dots
y_{m}$ and $y_{m+1}\dots y_{m+n}$ are not foreign sequences. Since
these two parts in the foreign sequence provide no additional
information in comparison with the MFS in them
\cite{Tan02Undermine}, they will be filtered out before further
analysis.

Fortunately, in the generation of the foreign sequence graph
(Algorithm \ref{alg:FSL}), the prefix sequence $y_{1}\dots y_{m}$
is filtered out automatically. Therefore, to collect the minimum
foreign sequence, only the suffix sequence $y_{m+1}\dots y_{m+n}$
needs to be eliminated. The method to filter out the suffix
sequence is trivial: if $FSL(e_{i})=FSL(e_{i-1})+1$, the foreign
sequence identified by $FSL(e_{i})$ will be filtered out since it
is included in the foreign sequence identified by $FSL(e_{i-1})$.

As stated before, the false alarms in the intrusive dataset or in
the test dataset can be identified and analyzed as well by the
intrusion context identification scheme. In summary, the scheme
will be useful to study the characteristics of the intrusions, to
remove the false alarms in the detection phase, and to improve the
efficiency of the AID detection techniques. However, every coin
has two sides. The identified intrusion context can be utilized to
design smarter intrusions, such as the information hiding
techniques \cite{Tan02HidingIntrusion} and the mimicry attacks
\cite{Wagner02MimicryAttacks}.

\subsection{Experimental evaluations}
In the experiments, the following aspects about the scheme will be
evaluated:
\begin{enumerate}
\item [A)] The effectiveness of the foreign sequence graph, and
how to identify the intrusion context;

\item [B)] The significance of the minimum foreign sequences in
the intrusive dataset.
\end{enumerate}

\subsubsection {Identifying the intrusion context from FSGs}
The following figures (Figure~\ref{fig:FSG}) summarize the foreign
sequence graphs for every intrusion into the chosen processes. For
the convenience of comparison, some FSG graphs are compressed into
one subfigure, and their borders are split with vertical lines for
different intrusive datasets. To easily identify the boundaries,
we introduce dummy values of $FSL=-4$ between different intrusive
datasets, and $FSL=-1$ between different processes in one dataset.

\begin{figure*}[t]
\centering \subfigure[The Process `named', and the Intrusion is
buffer overflow.] {\epsfig{file=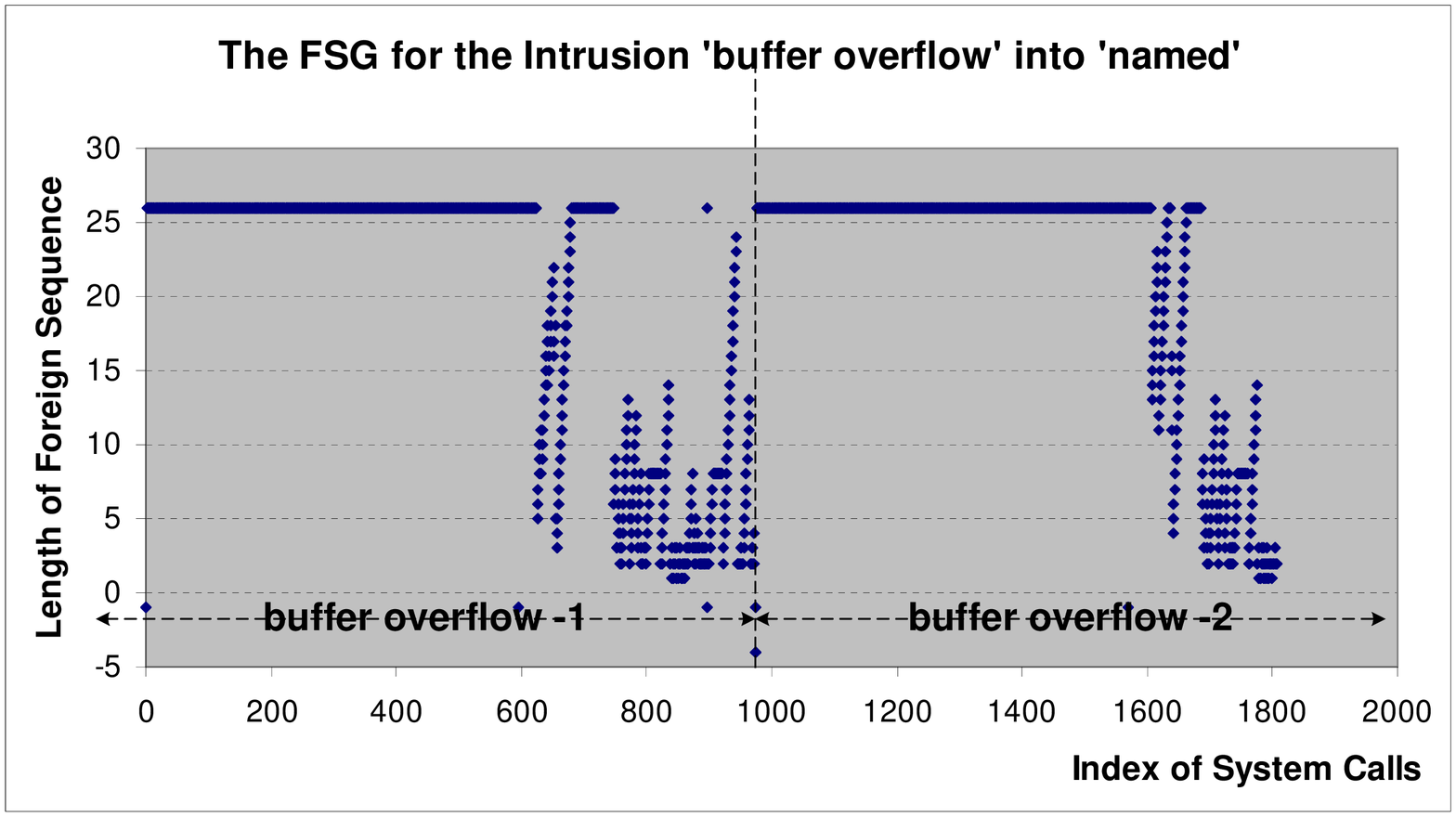, height=1.2in,
width=2.1in}} \subfigure[The Process `lpr', and the Intrusion is
lprcp.] {\epsfig{file=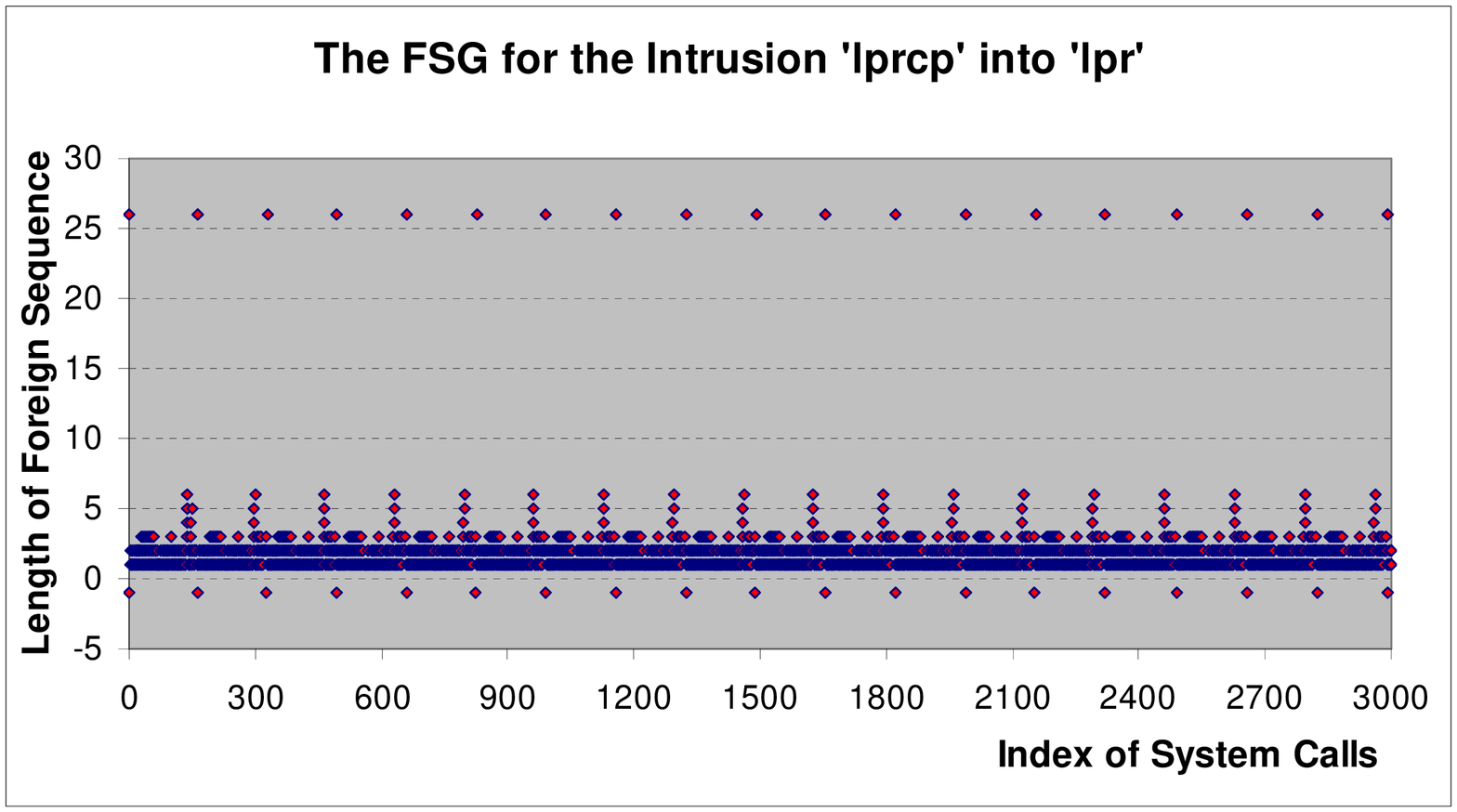, height=1.2in, width=2.1in}}
\subfigure[The Process `sendmail' from CERT, and the Intrusion is
syslog.] {\epsfig{file=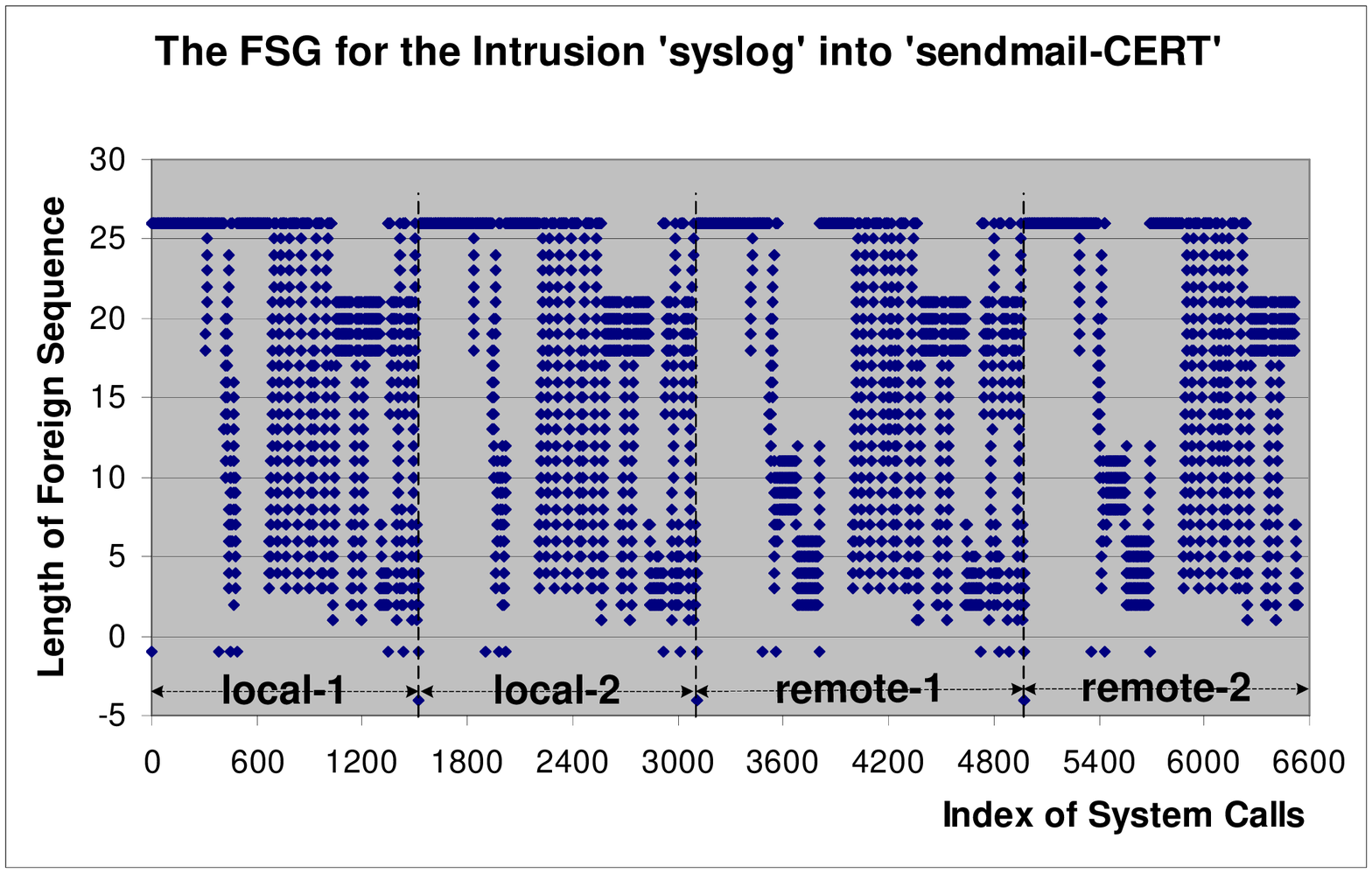, height=1.2in,
width=2.1in}} \subfigure[The Process `sendmail' from CERT, and the
Intrusions are sm565a and sm5x.]
{\epsfig{file=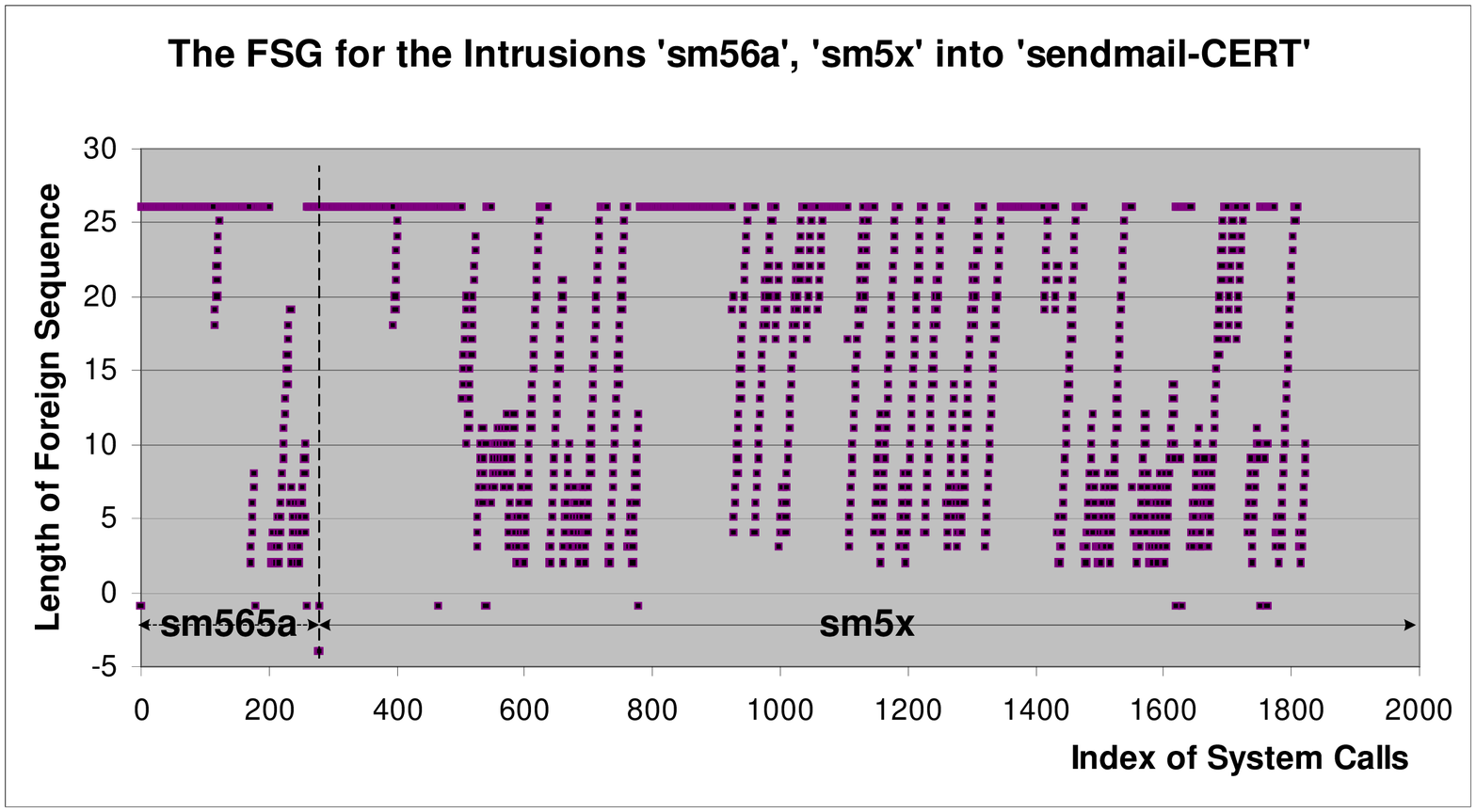, height=1.2in, width=2.1in}}
\subfigure[The Process `sendmail' from UNM, and the Intrusion is
decode.] {\epsfig{file=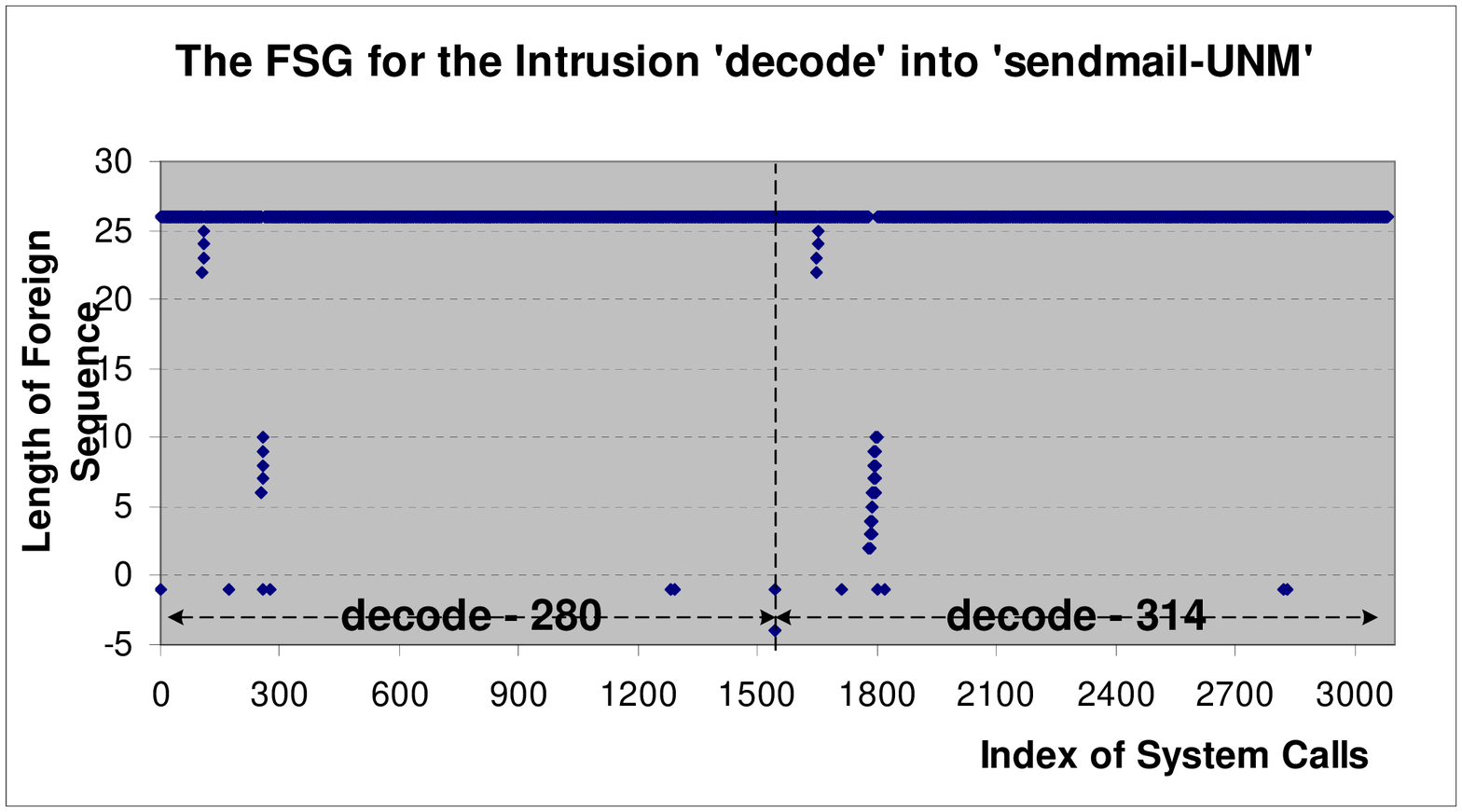, height=1.2in,
width=2.1in}} \subfigure[The Process `sendmail' from UNM, and the
Intrusion is forward loops.]
{\epsfig{file=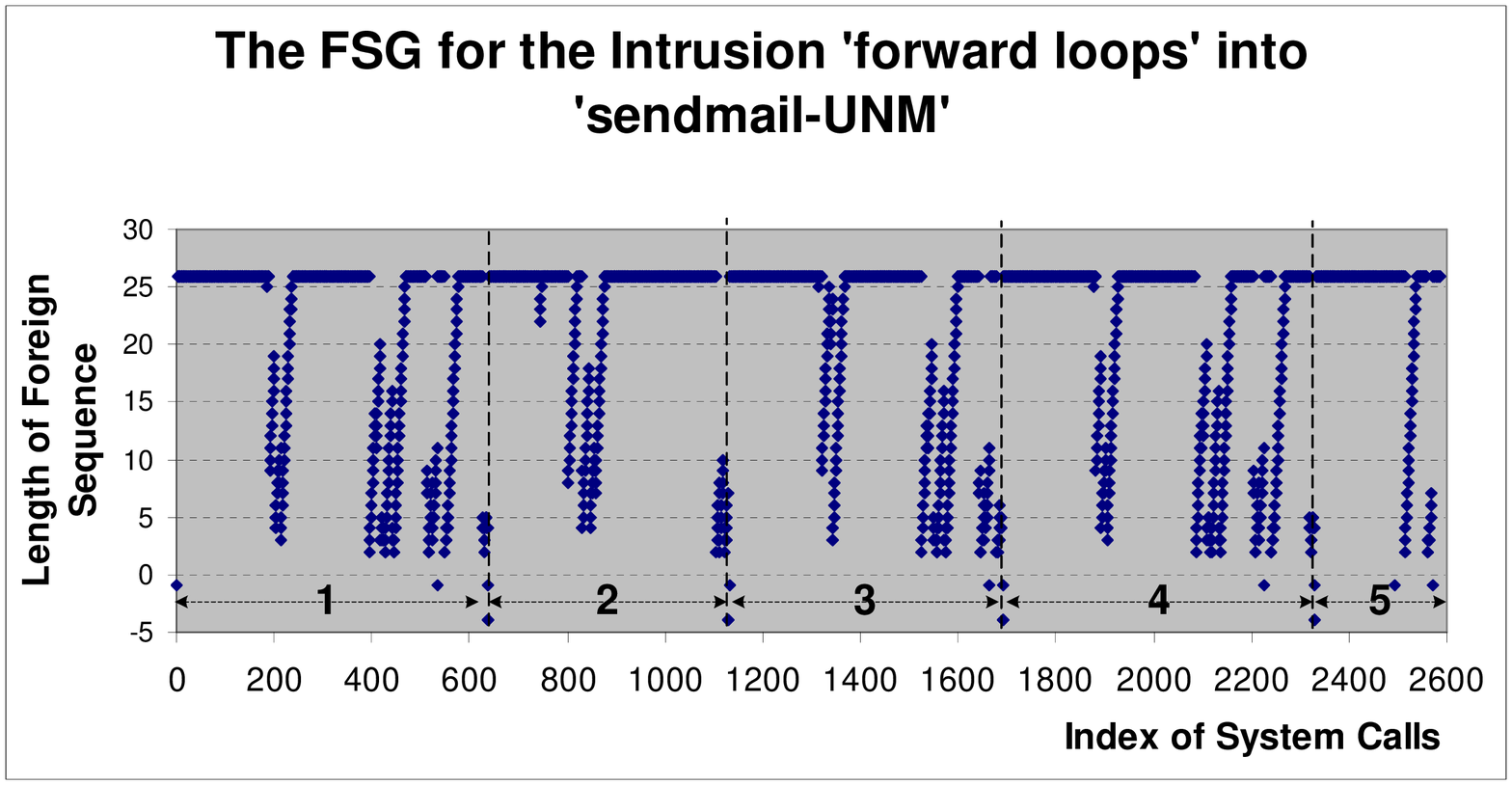, height=1.2in, width=2.1in}}
\subfigure[The Process `sendmail' from UNM, and the Intrusion is
sunsendmailcp (3 processes).] {\epsfig{file=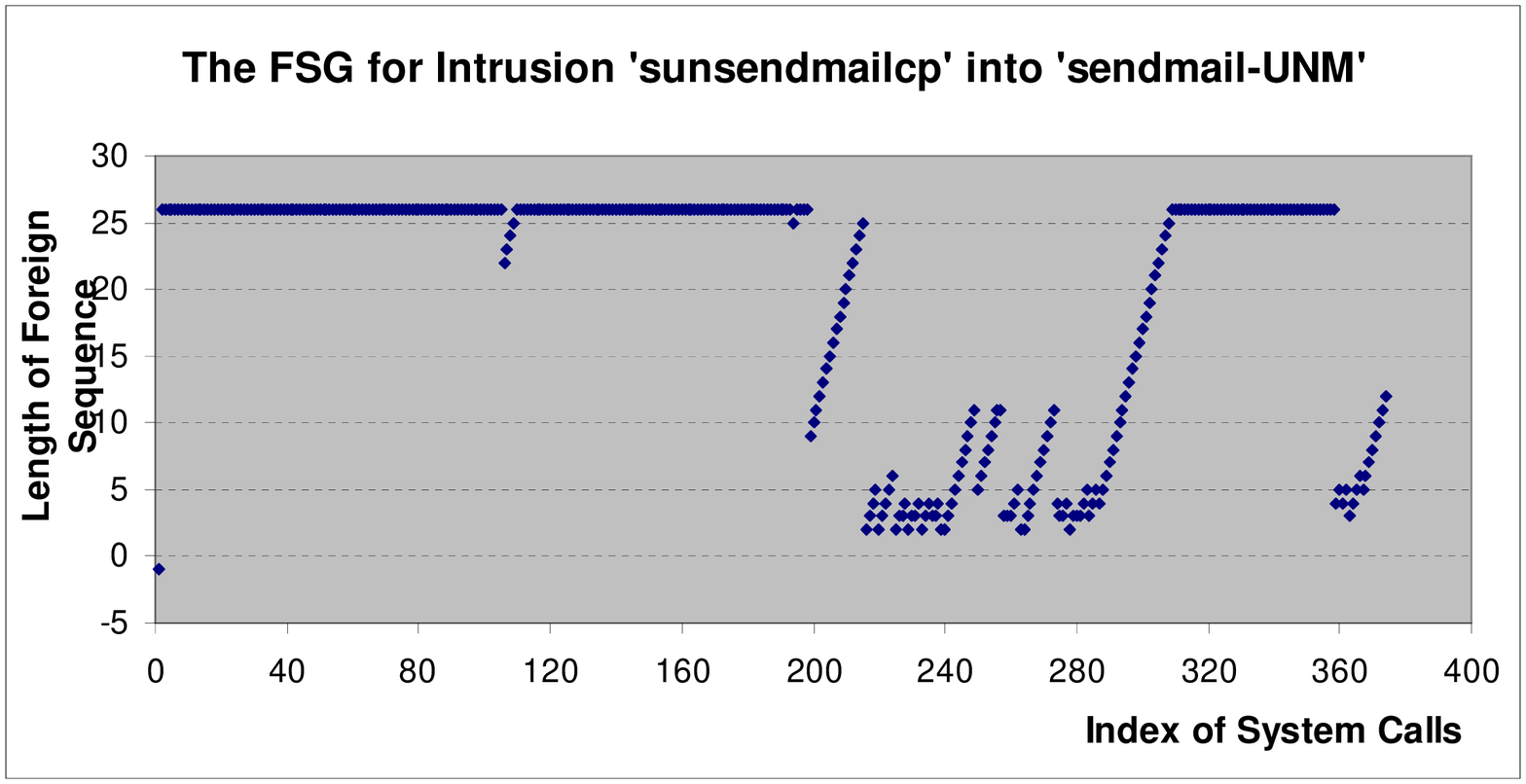,
height=1.2in, width=2.1in}} \subfigure[The Process `ftpd', and the
Intrusion is misconfiguration.] {\epsfig{file=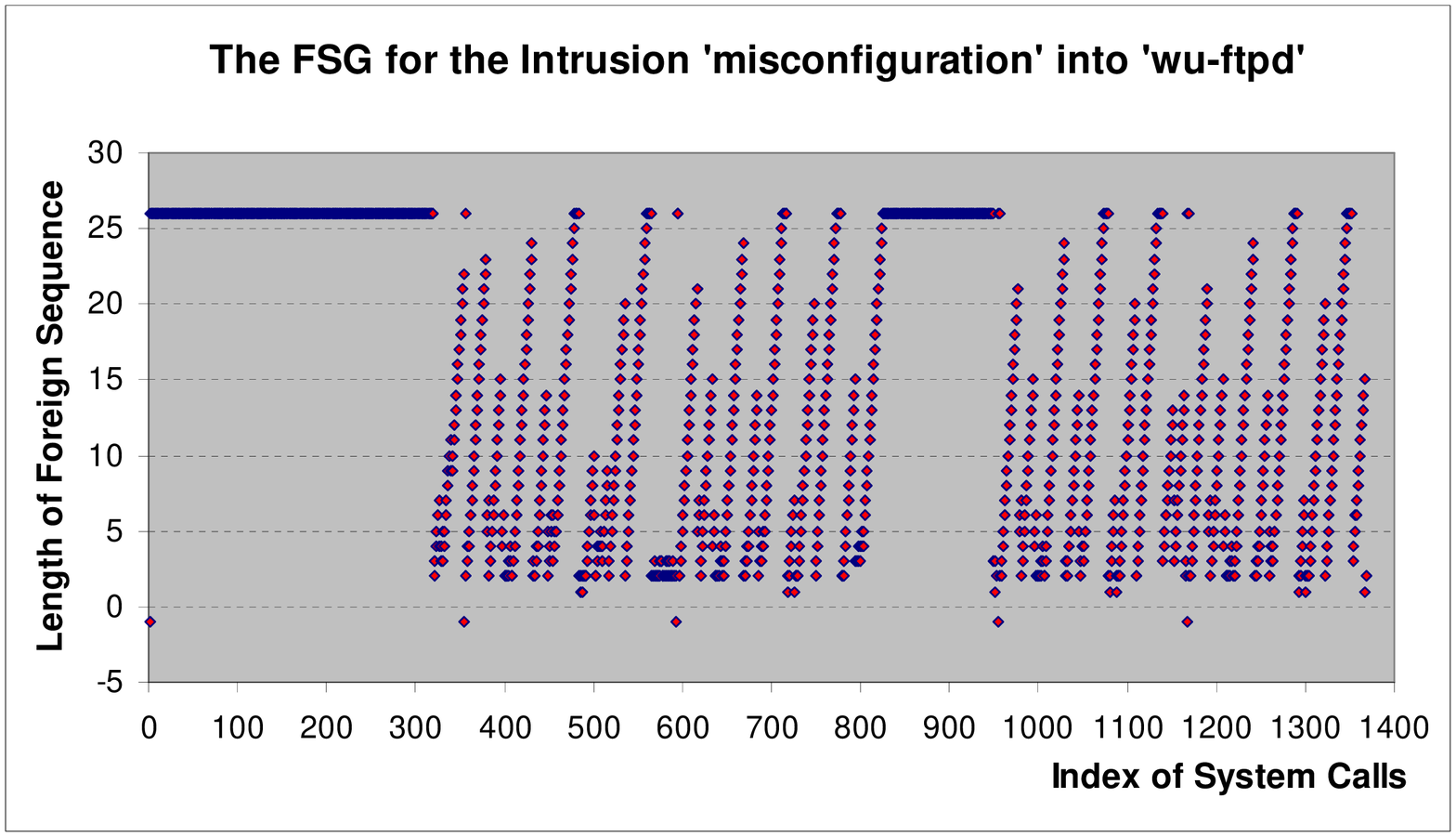,
height=1.2in, width=2.1in}} \subfigure[The Process `xlock', and
the Intrusion is buffer overflow.] {\epsfig{file=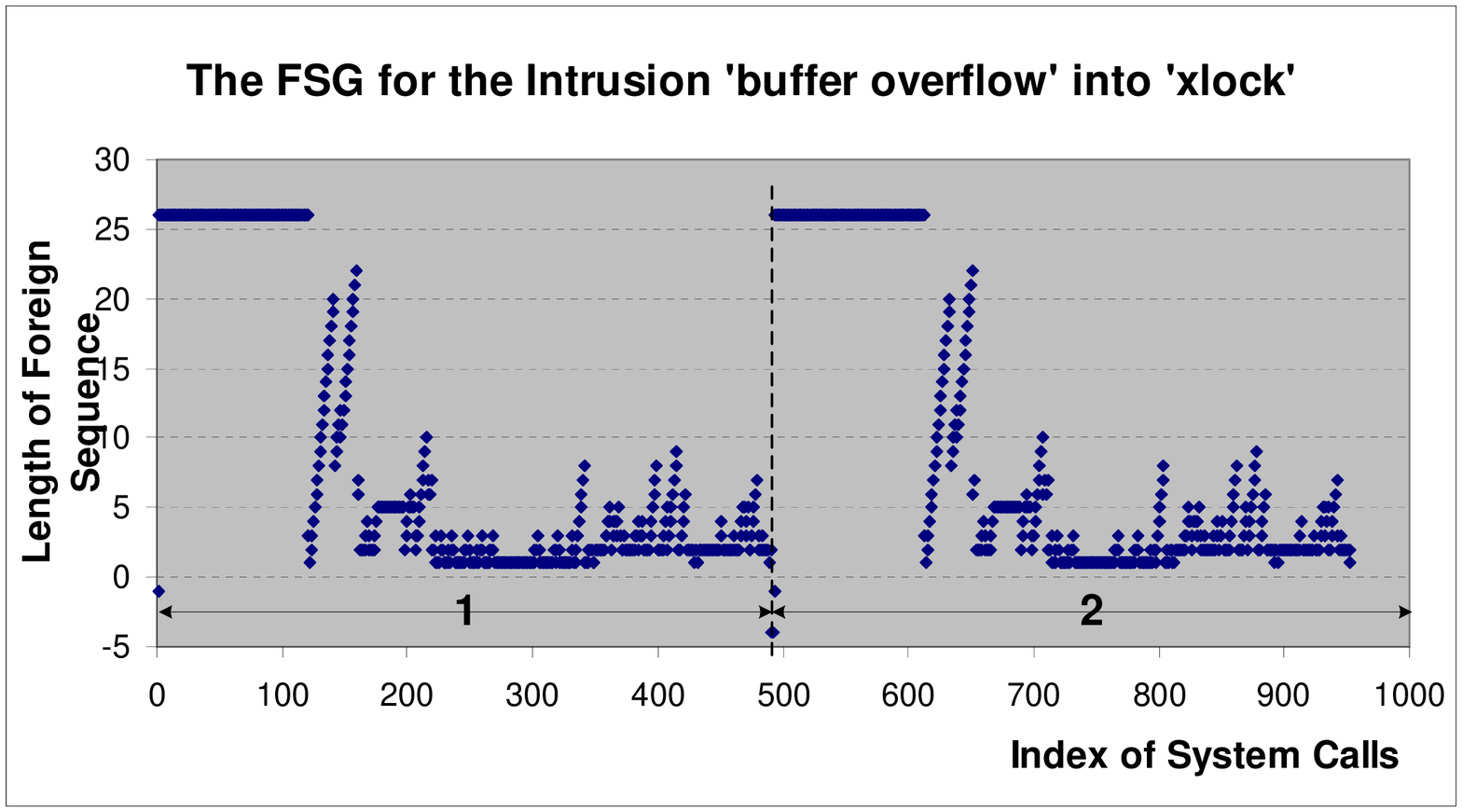,
height=1.2in, width=2.1in}} \caption{Foreign sequence graphs for
different intrusions and processes.} \label{fig:FSG}
\end{figure*}

From these foreign sequence graphs, we can make the following
observations:
\begin{itemize}
\item Some intrusions cannot be detected in the first (beginning)
stage by stide-like anomaly detectors since there are no foreign
sequences in that stage, such as the buffer overflow into `named';
yet some intrusions can be detected in the first stage, such as
the `lprcp' into `lpr from MIT';

\item Different runs of the same attack (intrusion) have almost
the same foreign sequence graphs (or intrusive characteristics),
such as the sunsendmailcp in which the three different runs
(10763, 10801, 10814) have the same foreign sequence graph;

\item For different intrusions into one process, the foreign
sequence graphs are not the same, and they are intrusion-specific.
Therefore, to detect all intrusions into a resource, one specific
detection strategy (such as the stide detector with defined
length) is not enough;

\item For most of the intrusions, there are obvious precursors at
the beginning of the anomalous events, and some of them are
manifested by the foreign sequences with larger length. It hints
at the existence of a tradeoff between the MMTA (Mean Time To
Alarm) of anomaly detectors and their efficiency (reflected by the
length of sequences).
\end{itemize}

\subsubsection{Minimum foreign sequences}
The minimum foreign sequences for `decode' are listed below. From
a look at the list, the answer to the `\textit{why 6?}' problem is
obvious (1:exit, 2:fork, 5:open, 6:close, 19:lseek, 95:connect,
112:vtrace).
\begin{itemize}
\item \textbf{decode-280} \\
process 283: 2-95-6-6-95-5
\item \textbf{decode-314} \\
process 317: 112-6, 6-19, 2-95-1, 2-95-6-6-95-5.
\end{itemize}

\begin{figure}[h]
\centering \epsfig{file=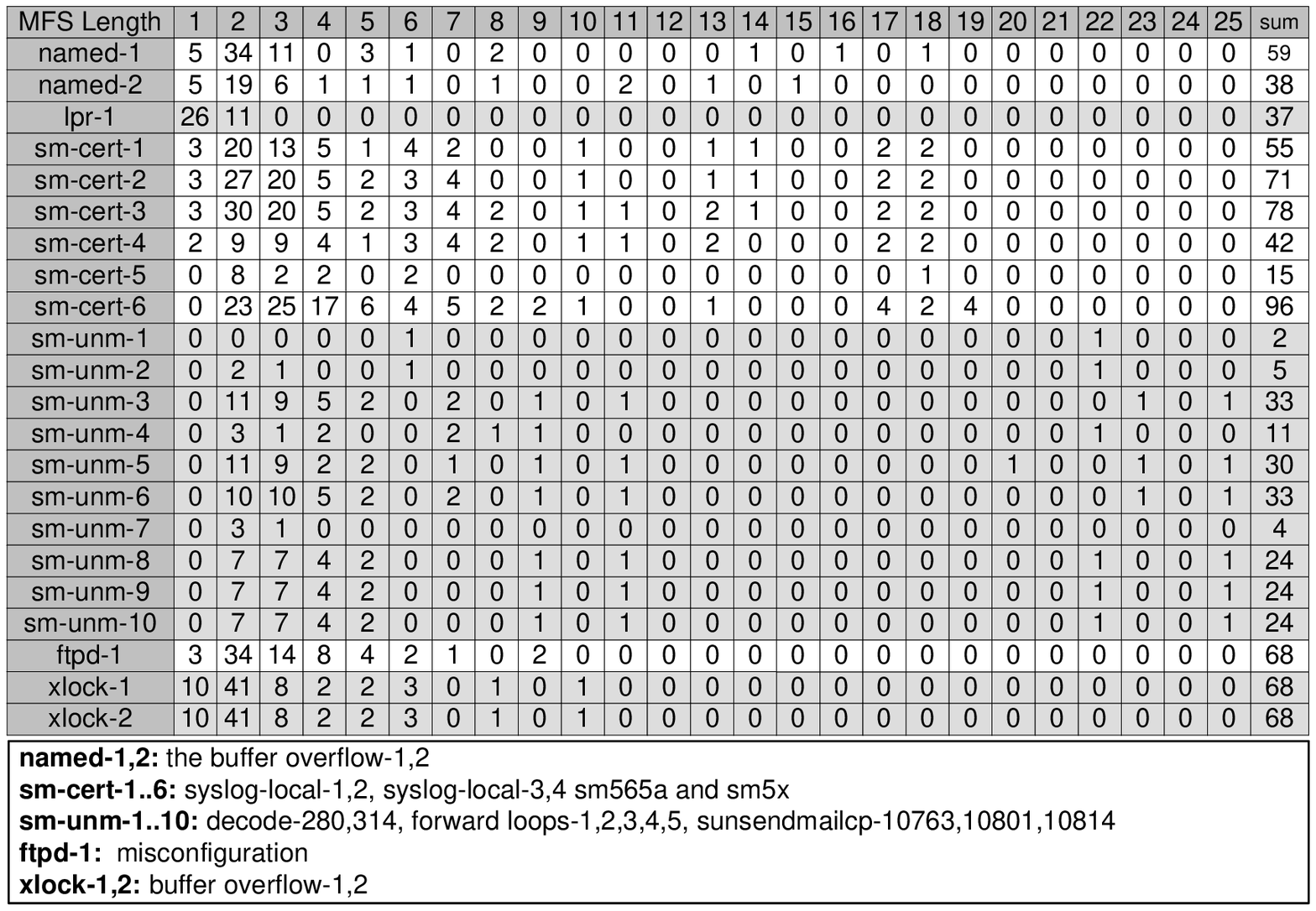,width=3.2in}
\caption{The number of Minimum Foreign Sequences (non-duplicated)
in all intrusive datasets.} \label{fig:numMFS}
\end{figure}

From the total number of non-duplicated MFSs in every intrusive
dataset (Figure~\ref{fig:numMFS}), we know that (1) the detector
window $DW=2$ is enough to detect most of the intrusions; (2) the
intrusion `\textit{decode-280}' leads to the magic number 6 for
stide; (3) if the detection window $DW>7$, the efficiency of stide
detectors will not be improved much as expected.

\begin{table}[h]
  \centering
  \caption{Shared Minimum Foreign Sequences by the same intrusion into the same process.}
  \label{tbl:sharedMFSs}
  \begin{scriptsize}
  \begin{tabular}{|l|c|c|}
    \hline
     Intrusion  &No. of MFSs    & No. of\\
                &of Each Run     &Shared MFSs\\
    \hline\hline decode  &\{2,5\}& 2 \\
    \hline buffer overflow into xlock &\{68,68\}& 68 \\
    \hline buffer overflow into named &\{59,38\}& 33 \\
    \hline sunsendmailcp     &\{24,24,24\}& 24 \\
    \hline forward loops  &\{33,11,30,33,4\}& 0 \\
    \hline syslog-local    &\{55,71\}& 52 \\
    \hline syslog-remote    &\{78,42\}& 42 \\
    \hline
  \end{tabular}
  \end{scriptsize}
\end{table}

In addition, we also note that different runs of the same
intrusion into one resource will share most of the minimum foreign
sequences (Table~\ref{tbl:sharedMFSs}). It is quite notable that
different runs of `sunsendmailcp' have the same set of minimum
foreign sequences. This discovery benefits the research on
anomaly-based intrusion detection because the diversity of
different runs of the same intrusion is not too large to design
one specific (or ad-hoc) IDS system for each of its runs. At the
same time, it strengthens two assertions that the foreign sequence
graphs are intrusion-specific, and that different runs of the same
intrusion have almost the same characteristics.

Also, the large quantity of minimum foreign sequences given in
Table~\ref{tbl:sharedMFSs}, especially for the intrusion `buffer
overflow', will discourage the mimicry attacks
\cite{Wagner02MimicryAttacks} and the information hiding paradigm
greatly. This is because all the large quantity of MFSs must be
mimicked to achieve a successful mimicry attack. In addition,
after a careful manual identification, the minimum foreign
sequences can also be applied to construct the `\textit{intrusion
signatures}' to be applied in signature-based intrusion detection
techniques.

\section{Conclusions and future work}
In this paper, a general framework is proposed to determine the
operational limits of stide detectors. Tan and Maxion
\cite{Kymie03DetermineOpLmts} in their attempt to solve the ``Why
six?"  problem, identified the length of the minimum foreign
sequence in the audit data as a lower bound for the length of
stide detectors. Our work complements their effort by showing the
effect of completeness of the normal model on stide's performance,
and establishing an upper bound for the length of the detector. In
addition to generalizing Tan and Maxion's results, this framework
provides a formal ground for analyzing future stide-like AID
detectors that are based on sequence analysis, by exploring the
dynamics of the various factors affecting operational limits of
stide, i.e. the false positives and true positives. Based on the
formal framework, the foundations of several work related to stide
are interpreted in a logical way.

The experiments we conducted not only validate our theoretical
results, they also provide further insights by clearly showing the
inter-dependencies of the various factors affecting stide's
performance, i.e., the influence of completeness of the training
dataset on stide efficiency is evaluated. The conclusion on the
completeness evaluation is that stide is not appropriate for
dynamic scenarios, such as the traffic behaviors in Internet.
Then, two applications of our framework are also designed to
demonstrate the usefulness of our framework. One is the trimming
procedure for the normal dataset, in which the redundant parts in
the normal dataset are filtered out for further analysis. To
achieve them, two graphical tools are designed to identify the
influence and the most compact critical section: MFS-MSS Average
Curve (MMAC) and MFS-MSS matrix (MMM).

From the MMAC curves, the influence of the completeness of the
training dataset on the MSSs in the test dataset and the MFSs in
the intrusive dataset are analyzed. The existence of the minimum
common false positive sequences are also confirmed in the MMAC
curves. At a finer granularity, the MMM matrix is utilized to find
the most compact critical section within the normal dataset for a
specific detection performance $\lambda$. The MMAC curves and the
MMM matrix also provide an intuitive indication of the complexity
of the corresponding process.

In this framework, the questions related to the `\textit{Why 6?}'
problem can be answered clearly, such as the question in
\cite{Kymie03DetermineOpLmts}, `\textit{to what extent can we
establish a link between detectable anomalies and intrusive
behaviors?}', and the answer lies in
Theorem~\ref{theorem:Tendency}. After analyzing the influence of
the completeness of the training dataset of a process on the
efficiency of the stide detectors, we can determine whether stide
is appropriate for detecting any intrusion into that process.

The second application of the framework, which is first introduced
here, is the intrusion context identification in an intrusive
dataset using the foreign sequence graphs. From the minimum
foreign sequences of intrusions, the following findings, which
will benefit the research on anomaly-based intrusion detection,
are reported:
\begin{enumerate}
\item Different runs of an intrusion almost have the same
characteristics; \item Different intrusions into one process will
cause different anomalies in the intrusive datasets; \item Most of
the intrusions have precursors, which are useful to provide short
MMTA; \item Some intrusions can not be detected in the first stage
when no anomalies are caused. \item There is diminishing rate of
return in terms of efficiency with the increase of the detector
window size.
\end{enumerate}

\textit{Limitations of the Framework.}
However, while using the proposed framework, we should bear in
mind certain limitations of the framework. These are briefly
stated below.
\begin{enumerate}
\item Like any AID technique, the framework is based on the
assumption that any anomaly in the intrusive dataset is an
indication of an intrusion into the resource. Even though the
assumption is reasonable under most circumstances, it is possible
that a non-malicious access that deviates from the normal behavior
will be detected as an intrusion. On the other hand, if an
intrusion can successfully mimic the normal behaviors, then no AID
technique can detect such an intrusion\cite{Tan02HidingIntrusion}.

\item The framework specifically deals with stide-like AID
techniques which assumes that intrusions are manifested in the
sequences of system calls and tries to detect them by a systematic
analysis of such sequences.  Therefore it may not be appropriate
to apply it to all possible intrusions or detection techniques
\cite{Wagner02MimicryAttacks} \cite{Tan02HidingIntrusion}.
\end{enumerate}

In our future work, the framework will be further evaluated by the
datasets under different environments, e.g. the networks and the
windows platform. Then, the definitions for effective, complete
and efficient anomaly-based intrusion detectors will be
generalized to other sequence-based AID techniques. As a practical
and promising method to analyze the intrusion characteristics, the
mechanisms for intrusion context identification will also be
extended to other AID techniques in our further study.

\section{Acknowledgements}
We are grateful to Prof. Forrest and her research group at the
University of New Mexico, for their scientific generosity, since
the authors downloaded their datasets and their documents from
\cite{Dataset94UNM}.

\bibliographystyle{ieee}
\bibliography{all}
\newpage
\onecolumn
\appendix
\section{Proofs of some Theorems}
\subsection{Proof of Theorem~\ref{theorem:MFS=MSS+1}}
\begin{proof}
Assume that $S\in MFS_{min}(\Sigma_{tgt}|\Sigma_{ref})$. Thus,
$|S| =|MFS|_{min}(\Sigma_{tgt}|\Sigma_{ref})$. From the definition
of $MFS(\Sigma_{tgt}|\Sigma_{ref})$, $S\in
FRGN(\Sigma_{tgt}|\Sigma_{ref})$, and there exists an 1-order
subsequence $S'\in SELF(\Sigma_{tgt}|\Sigma_{ref})$ at least
considering that $FRGN(\Sigma_{tgt}|\Sigma_{ref})\cup
SELF(\Sigma_{tgt}|\Sigma_{ref})=SS(\Sigma_{tgt})$. Therefore,
$(S'\in SELF(\Sigma_{tgt}|\Sigma_{ref}))\wedge(S\in
SS(\Sigma_{tgt}))\wedge(S\succcurlyeq_{1} S')\wedge(S\not\in
SELF(\Sigma_{tgt}|\Sigma_{ref}))=True$, and we can conclude that
$S'\in MSS(\Sigma_{tgt}|\Sigma_{ref})$.

Then, using proof by contradiction to prove that $S'\in
MSS_{min}(\Sigma_{tgt}|\Sigma_{ref})$. If $S'\not\in
MSS_{min}(\Sigma_{tgt}|\Sigma_{ref})$, there must be another
sequence $S''\in MSS_{min}(\Sigma_{tgt}|\Sigma_{ref})$, and
$|S''|<|S'|$. Following above deduction, we can determine that
there is one $S'''\in MFS(\Sigma_{tgt}|\Sigma_{ref})$, and
$S'''\succcurlyeq_{1} S''$. Considering $|S|=|S'|+1$ and
$|S'''|=|S''|+1$, we get $|S'''|<|S|$, which contradicts to $|S|
=|MFS|_{min}(\Sigma_{tgt}|\Sigma_{ref})$ as $S'''\in
MFS(\Sigma_{tgt}|\Sigma_{ref})$.

Considering that $S\in MFS_{min}(\Sigma_{tgt}|\Sigma_{ref})$,
$S'\in MSS_{min}(\Sigma_{tgt}|\Sigma_{ref})$, and $|S|=|S'|+1$,
the following equation is held:
\begin{eqnarray}
|MSS|_{min}(\Sigma_{tgt}|\Sigma_{ref})=|MFS|_{min}(\Sigma_{tgt}|\Sigma_{ref})-1\nonumber
\end{eqnarray}
\end{proof}

\subsection{Proof of Theorem~\ref{theorem:effective}}
\begin{proof}
From the definitions, we know that:
\begin{eqnarray*}
|MFS|_{min}(\Sigma_{int}|\Sigma_{trn}) =\min(\{l|l>0;
SS(\Sigma_{int}, l)-SS(\Sigma_{trn}, l)\neq\Phi\})
\end{eqnarray*}
(\textbf{1}) If $\omega\geqslant
|MFS|_{min}(\Sigma_{int}|\Sigma_{trn})$, then,
\begin{eqnarray*}
SS(\Sigma_{int}, \omega)-SS(\Sigma_{trn}, \omega)\neq\Phi
\Longrightarrow TPSS(\Sigma_{int}|\Sigma_{trn}, \omega)\neq \Phi
\end{eqnarray*}
Hence, the stide detector of length $\omega$ built from
$\Sigma_{trn}$ is effective w.r.t. $\Sigma_{int}$.

\noindent (\textbf{2}) If the stide detector with the length
$\omega$ built by $\Sigma_{trn}$ is effective w.r.t.
$\Sigma_{int}$, then,
\begin{eqnarray*}
TPSS(\Sigma_{int}|\Sigma_{trn}, \omega)\neq \Phi &\Longrightarrow&
SS(\Sigma_{int}, \omega)-SS(\Sigma_{trn},
\omega)\neq\Phi\\
&\Longrightarrow& \omega\geqslant
|MFS|_{min}(\Sigma_{int}|\Sigma_{trn})
\end{eqnarray*}
\end{proof}

\subsection{Proof of Theorem~\ref{theorem:complete}}
\begin{proof}
From the definitions, we know that:
\begin{eqnarray*}
|MSS|_{min}(\Sigma_{tst}|\Sigma_{trn})=\max(\{l|l\geqslant 0;
SS(\Sigma_{tst}, l)-SS(\Sigma_{trn}, l)=\Phi\})
\end{eqnarray*}
(\textbf{1}) If $\omega\leq
|MSS|_{min}(\Sigma_{tst}|\Sigma_{trn})$, then,
\begin{eqnarray*}
SS(\Sigma_{tst}, \omega)-SS(\Sigma_{trn}, \omega)=\Phi
\Longrightarrow FPSS(\Sigma_{tst}|\Sigma_{trn}, \omega)= \Phi
\end{eqnarray*}
Hence, the stide detector with the length $\omega$ built by
$\Sigma_{trn}$ is complete w.r.t. $\Sigma_{tst}$.

\noindent (\textbf{2}) If the stide detector with the length
$\omega$ built by $\Sigma_{trn}$ is complete w.r.t.
$\Sigma_{tst}$, then,
\begin{eqnarray*}
FPSS(\Sigma_{tst}|\Sigma_{trn}, \omega)= \Phi &\Longrightarrow &
SS(\Sigma_{tst}, \omega)-SS(\Sigma_{trn},
\omega)=\Phi\\
&\Longrightarrow & \omega\leq
|MSS|_{min}(\Sigma_{tst}|\Sigma_{trn})
\end{eqnarray*}
\end{proof}

\subsection{Proof of Theorem~\ref{theorem:minMFPS-MFS}}
\begin{proof}
From the MFS definition,
\begin{eqnarray*}
|MFS|_{min}(\Sigma_{int}|\Sigma_{trn})=min(\{l|l>0;SS(\Sigma_{int},
l)-SS(\Sigma_{trn}, l)\neq\Phi\})
\end{eqnarray*}
The foreign sequence length vector for $\Sigma_{trn}$ and
$\Sigma_{int}$ is
\begin{eqnarray*}
FSLV(\Sigma_{int}|\Sigma_{trn})=\{l|l\geqslant0;\;SS(\Sigma_{int},l)-SS(\Sigma_{trn},
l)\neq\Phi\}
\end{eqnarray*}
For $\forall l\in FSLV(\Sigma_{int}|\Sigma_{trn})$,
\begin{eqnarray*}
\lefteqn{SS(\Sigma_{int}, l)-SS(\Sigma_{trn}, l)\neq\Phi}\\
&\Leftrightarrow&\exists S(S\in (SS(\Sigma_{int}, l)-SS(\Sigma_{trn}, l)))\\
&\Leftrightarrow& \exists S(S\in SS(\Sigma_{int}, l)\wedge
S\not\in
SS(\Sigma_{trn}, l))\\
&\Leftrightarrow& \exists S(S\in SS(\Sigma_{int}, l)\wedge
S\not\in SS(\Sigma_{trn}, l)
\wedge(S\in SS(\Sigma_{tst}, l)\vee S\not\in SS(\Sigma_{tst}, l)))\\
&\Leftrightarrow& \exists S((S\in SS(\Sigma_{int}, l)\wedge
S\not\in SS(\Sigma_{trn}, l) \wedge S\in SS(\Sigma_{tst}, l))\vee
(S\in SS(\Sigma_{int}, l)\wedge S\not\in SS(\Sigma_{trn}, l)\wedge
S\not\in SS(\Sigma_{tst}, l)))\\
&\Leftrightarrow& \exists S((S\in SS(\Sigma_{int}, l)\wedge
S\not\in SS(\Sigma_{trn}, l)\wedge S\in SS(\Sigma_{tst}, l))\vee
(S\in SS(\Sigma_{int},l)\wedge S\not\in (SS(\Sigma_{trn}, l)\vee
SS(\Sigma_{tst}, l))))\\
&\Leftrightarrow& \exists S((S\in SS(\Sigma_{int}, l)\wedge S\in(
SS(\Sigma_{tst}, l)- SS(\Sigma_{trn}, l)))\vee (S\in
SS(\Sigma_{int}, l)\wedge S\not\in
SS(\Sigma_{trn}\odot\Sigma_{tst}, l)))\\
&\Leftrightarrow& \exists S((S\in SS(\Sigma_{int}, l)\wedge S\in
FPSS(\Sigma_{tst}|\Sigma_{trn}, l))\vee (S\in SS(\Sigma_{int},
l)\wedge S\not\in
SS(\Sigma_{trn}\odot\Sigma_{tst}, l)))\\
&\Leftrightarrow& \exists S(S\in (SS(\Sigma_{int}, l)\cap
FPSS(\Sigma_{tst}|\Sigma_{trn}, l))\vee S\in (SS(\Sigma_{int}, l)-
SS(\Sigma_{trn}\odot\Sigma_{tst}, l)))\\
&\Leftrightarrow& \exists S(S\in ((SS(\Sigma_{int}, l)\cap
FPSS(\Sigma_{tst}|\Sigma_{trn}, l))\cup (SS(\Sigma_{int}, l)- SS(\Sigma_{trn}\odot\Sigma_{tst}, l))))\\
&\Leftrightarrow& SS(\Sigma_{int}, l)\cap
FPSS(\Sigma_{tst}|\Sigma_{trn}, l)\neq\Phi\vee SS(\Sigma_{int},
l)- SS(\Sigma_{trn}\odot\Sigma_{tst}, l)\neq\Phi
\end{eqnarray*}
Hence,
\begin{eqnarray*}
FSLV(\Sigma_{int}|\Sigma_{trn}) &=&\{l|l\geqslant 0;
SS(\Sigma_{int}, l)\cap FPSS(\Sigma_{tst}|\Sigma_{trn},
l)\neq\Phi\}\cup\{l|l\geqslant 0;SS(\Sigma_{int}, l)-
SS(\Sigma_{trn}\odot\Sigma_{tst}, l)\neq\Phi\}
\end{eqnarray*}
Finally,
\begin{eqnarray*}
\lefteqn{|MFS|_{min}(\Sigma_{int}|\Sigma_{trn})}\\
&=&\min(FSLV(\Sigma_{int}|\Sigma_{trn}))\\
&=&\min(\{l|SS(\Sigma_{int}, l)\cap
FPSS(\Sigma_{tst}|\Sigma_{trn},
l)\neq\Phi\}\cup\{l|SS(\Sigma_{int}, l)-
SS(\Sigma_{trn}\odot\Sigma_{tst},
l)\neq\Phi\}) \\
&=&\min(\min(\{l|SS(\Sigma_{int}, l)\cap FPSS(\Sigma_{tst}|\Sigma_{trn}, l)\neq\Phi\}),\\
&&\min(\{l|SS(\Sigma_{int}, l)- SS(\Sigma_{trn}\odot\Sigma_{tst},
l)\neq\Phi\})) \\
&=&\min(|CFPS|_{min}(\Sigma_{int},\Sigma_{tst}|\Sigma_{trn}),|MFS|_{min}(\Sigma_{int}|\Sigma_{trn}\odot\Sigma_{tst}))
\end{eqnarray*}
\end{proof}

\subsection{Proof of Theorem~\ref{theorem:Tendency}}
\begin{proof}
If $|MSS|_{min}(\Sigma_{tst}|\Sigma_{trn})\geqslant 1$,
\begin{eqnarray}
\label{eqn:MSSnMFPS} \forall l(1\leq l\leq
|MSS|_{min}(\Sigma_{tst}|\Sigma_{trn}),
FPSS(\Sigma_{tst}|\Sigma_{trn}, l)=\Phi)
\end{eqnarray}
Furthermore, according to the definition of $CFPS$,
\begin{equation}
\label{eqn:MSSMFPSRelation} |CFPS|_{min}(\Sigma_{int},
\Sigma_{tst}|\Sigma_{trn})>|MSS|_{min}(\Sigma_{tst}|\Sigma_{trn})
\end{equation}
($\Longleftarrow$) If
$|MSS|_{min}(\Sigma_{tst}|\Sigma_{trn})\geqslant |MFS|_{min}(
\Sigma_{int}|\Sigma_{trn}\odot\Sigma_{tst})$, there exist
efficient stide detectors for datasets $\Sigma_{trn}$,
$\Sigma_{tst}$ and $\Sigma_{int}$.
\begin{eqnarray*}
\lefteqn{|MSS|_{min}(\Sigma_{tst}|\Sigma_{trn})\geqslant
|MFS|_{min}(\Sigma_{int}|\Sigma_{trn}\odot\Sigma_{tst})}\\
&\Rightarrow&|MSS|_{min}(\Sigma_{tst}|\Sigma_{trn})\geqslant 1\\
&\Rightarrow&|CFPS|_{min}(\Sigma_{int},
\Sigma_{tst}|\Sigma_{trn})>|MSS|_{min}(\Sigma_{tst}|\Sigma_{trn})\\
&\Rightarrow&|CFPS|_{min}(\Sigma_{int},
\Sigma_{tst}|\Sigma_{trn})>|MFS|_{min}(\Sigma_{int}|\Sigma_{trn}\odot\Sigma_{tst})\\
&\Rightarrow&|MFS|_{min}(\Sigma_{tst}|\Sigma_{trn})=|MFS|_{min}(\Sigma_{int}|\Sigma_{trn}\odot\Sigma_{tst})\\
&\Rightarrow&|MFS|_{min}(\Sigma_{tst}|\Sigma_{trn})\leq
|MSS|_{min}(\Sigma_{tst}|\Sigma_{trn})
\end{eqnarray*}
From Theorem~\ref{theorem:MSS-MFSefficient}, there exist efficient
stide detectors under this scenario.

\noindent($\Longrightarrow$) If there exist efficient stide
detectors for $\Sigma_{trn}$, $\Sigma_{tst}$ and $\Sigma_{int}$,
$|MSS|_{min}(\Sigma_{tst}|\Sigma_{trn})\geqslant
|MFS|_{min}(\Sigma_{int}|\Sigma_{trn}\odot\Sigma_{tst})$.

\noindent To apply the proof by contradiction, let us assume
$|MSS|_{min}(\Sigma_{tst}|\Sigma_{trn})<
|MFS|_{min}(\Sigma_{int}|\Sigma_{trn}\odot\Sigma_{tst})$.

\noindent(\S 1) If $|MSS|_{min}(\Sigma_{tst}|\Sigma_{trn})=0$,\\
From the MFS definition,
$|MFS|_{min}(\Sigma_{tst}|\Sigma_{trn})\geqslant 1 >
|MSS|_{min}(\Sigma_{tst}|\Sigma_{trn})$. Based on
Theorem~\ref{theorem:MSS-MFSefficient}, there does not exist
efficient stide detectors, and that is contradict with our
statement. So, $|MSS|_{min}(\Sigma_{tst}|\Sigma_{trn})<
|MFS|_{min}(\Sigma_{int}|\Sigma_{trn}\odot\Sigma_{tst})$ is not
correct.

\noindent(\S 2) If $|MSS|_{min}(\Sigma_{tst}|\Sigma_{trn})\geqslant 1$,\\
(2.a) If $|CFPS|_{min}(\Sigma_{int},
\Sigma_{tst}|\Sigma_{trn})\geqslant
|MFS|_{min}(\Sigma_{int}|\Sigma_{trn}\odot\Sigma_{tst})$,\\
From equation (\ref{eqn:IncompleteMFS}), $
|MFS|_{min}(\Sigma_{int}|\Sigma_{trn}) =
|MFS|_{min}(\Sigma_{int}|\Sigma_{trn}\odot\Sigma_{tst}) $.
Furthermore, we assume
$|MFS|_{min}(\Sigma_{int}|\Sigma_{trn}\odot\Sigma_{tst})>|MSS|_{min}(\Sigma_{tst}|\Sigma_{trn})$,
therefore
\begin{eqnarray*}
|MFS|_{min}(\Sigma_{int}|\Sigma_{trn})>|MSS|_{min}(\Sigma_{tst}|\Sigma_{trn})
\end{eqnarray*}
\noindent(2.b) If $|CFPS|_{min}(\Sigma_{int},
\Sigma_{tst}|\Sigma_{trn})<
|MFS|_{min}(\Sigma_{int}|\Sigma_{trn}\odot\Sigma_{tst})$,\\
From Eqn.\ref{eqn:IncompleteMFS}, $
|MFS|_{min}(\Sigma_{int}|\Sigma_{trn})=|CFPS|_{min}(\Sigma_{int},
\Sigma_{tst}|\Sigma_{trn}) $, and Eqn.(\ref{eqn:MSSMFPSRelation}),
\begin{eqnarray*}
|CFPS|_{min}(\Sigma_{int},
\Sigma_{tst}|\Sigma_{trn})>|MSS|_{min}(\Sigma_{tst}|\Sigma_{trn})\\
\Rightarrow|MFS|_{min}(\Sigma_{int}|\Sigma_{trn})>|MSS|_{min}(\Sigma_{tst}|\Sigma_{trn}).
\end{eqnarray*}
From (2.a), (2.b) and Theorem~\ref{theorem:MSS-MFSefficient},
there are no efficient stide detectors, and that is contradict to
the statement. Therefore, $|MSS|_{min}(\Sigma_{tst}|\Sigma_{trn})<
|MFS|_{min}(\Sigma_{int}|\Sigma_{trn}\odot\Sigma_{tst})$ is not
correct.

Based on (1) and (2), the theorem is proved.
\end{proof}

\end{document}